%% file: alagi_NRCLfull.tex
\documentclass[a4paper]{article}
\input{finite-style/mystyle-core}
\usepackage{authblk}

\begin{document}
\title{NRCL - A Model Building Approach to\\ the {\mEPR} Fragment\\ (Full Paper)}
\setcounter{Maxaffil}{3}
\author[1,2,3]{G\'abor Alagi}
\author[3]{Christoph Weidenbach}
\affil[1]{
	Saarbr\"ucken Graduate School of Computer Science, Germany
}
\affil[2]{
	Saarland University, 
	Saarbr\"ucken, Germany
}
\affil[3]{
  Max-Planck-Institut f\"ur Informatik,
  Saarbr\"ucken, Germany
}
\affil[ ]{
	\textbf{\{galagi, weidenbach\}@mpi-inf.mpg.de}
}

\date{}

\clearpage

\maketitle

\begin{abstract}
We combine constrained literals for model representation 
with key concepts from first-order superposition
and propositional conflict-driven clause learning (CDCL)
to create the new calculus \emph{Non-Redundant Clause Learning} 
(NRCL) deciding the {\mEPR} fragment. 
Our calculus uses first-order literals constrained by disequations between tuples of terms 
for compact model representation.
From superposition, NRCL inherits the abstract redundancy criterion and 
the monotone model operator.
CDCL adds the dynamic, conflict-driven search for an 
atom ordering inducing a model. 
As a result, in NRCL a false clause can
be found effectively modulo the current model candidate. It guides
the derivation of a first-order ordered resolvent that is never redundant.
Similar to 1UIP-learning in CDCL, the learned resolvent induces
backtracking and, by blocking the previous conflict state via propagation, 
it enforces progress towards finding a model or a refutation.
The non-redundancy result also implies that only finitely many clauses
can be generated by NRCL on the {\mEPR} fragment, which 
serves as an argument for termination.
\end{abstract}
\input{src/s1-intro/introduction}
\input{src/s2-defs/definitions}
\input{src/s3-calc/calculus}
\input{src/s4-decproc/soundness}
\input{src/s4-decproc/regruns}
\input{src/s4-decproc/redundancy}
\input{src/s4-decproc/decision-procedure}
\input{src/s8-tech/technicalities}
\input{src/s6-relw/related-work}
\input{src/s7-futw/future-work}
\bibliographystyle{mpiabbrv-mod}
\bibliography{alagi_refs}
\ifdefined\PRINTNOTES
\input{notes/notes-main}
\fi
\end{document}

%% file: finite-style/mystyle-core.tex
\usepackage{paralist}
\usepackage{enumerate}

\usepackage{amsmath}
\usepackage{amssymb}
\usepackage{latexsym}
\usepackage{stmaryrd}

\newcommand{\mGnd}{\operatorname{gnd}} 
\newcommand{\mLVar}{\operatorname{lvar}} 
\newcommand{\mRVar}{\operatorname{rvar}} 
\newcommand{\mDmn}{\operatorname{dom}} 
\newcommand{\mRng}{\operatorname{rng}} 
\newcommand{\mMGU}{\operatorname{mgu}} 
\newcommand{\mDef}{\operatorname{def}} 
\newcommand{\mDomain}{\mathcal{D}} 
\newcommand{\mVar}{\operatorname{var}} 

\newcommand{\mEPR}{Bernays-Sch\"onfinkel}
\newcommand{\mEPRs}{BS}

\usepackage{alltt}
\usepackage[ruled, lined, linesnumbered]{algorithm2e}

\newcommand{\startproof}{{\bf Proof:~}}
\newcommand{\finishproof}{{\bf Qed.}}

\newcommand{\leaveabit}{\\[6 pt]}
\newtheorem{defi}{Definition}[section]
\newtheorem{remark}[defi]{Remark}
\newtheorem{theo}[defi]{Theorem}
\newtheorem{prop}[defi]{Proposition}
\newtheorem{coro}[defi]{Corollary}
\newtheorem{example}[defi]{Example}
\newtheorem{lemm}[defi]{Lemma}

\newcommand\exClauseNo[1]{C_{#1}} 
\newcommand{\rst}[1]{|_{#1}} 

\newcommand{\ie}{i.e.\ }
\newcommand{\eg}{e.g.\ }

\newcommand{\wrt}{w.r.t.\ }

%% file: src/s1-intro/introduction.tex
\section{Introduction}
The {\mEPR} fragment, also called Effectively Propositional Logic, or BS (or EPR) for short, 
is an important fragment of classic first-order logic, where only constants are allowed as
function symbols in the clause normal form. 

This decidable and NEXPTIME-complete fragment has many applications,
including knowledge representation~\cite{HustadtMS04} and ontological reasoning~\cite{SudaWeidenbachWischnewskiIJCAR10},
hardware verification~\cite{KhasidashviliKV09}\cite{PerezV07}\cite{EmmerKKV10}, logic programming~\cite{EiterFT05}, and planning~\cite{PerezV13}.

Over the years a number of calculi have attempted to provide an efficient 
solution for {\mEPRs} problems. 
These approaches range from the early \emph{SEM} and \emph{Mace} systems~\cite{Tammet03finitemodel} to the recent state-of-the-art solvers 
like \emph{iProver}~\cite{InstGen03} and \emph{Darwin}~\cite{ModelEvolution03}, but even general purpose first-order theorem provers 
provide specialized techniques for {\mEPRs} problems, 
like \emph{generalisation} in \emph{Vampire}~\cite{Generalization08}, or 
specialized splitting techniques for \emph{SPASS} introduced in~\cite{HillenbrandWeidenbach13} and \cite{FietzkeW09}. 

In this paper, we introduce a new calculus for solving {\mEPRs} problems with 
iterative model building.
Our approach builds first-order candidate models instead of approximations, uses 
constrained literals for model representation, and learns new non-redundant clauses to guide the search.

Our calculus, called NRCL or \emph{Non-Redundant Clause Learning}, 
shares many principles with propositional SAT solving and superposition.
For a detailed introduction to \emph{conflict-driven clause learning (CDCL)}, 
see the early article \cite{SilvaS96}, or the more recent handbook \cite{SATHandbook}.
The interested reader can get a thorough overview of superposition in the articles 
\cite{BachmairGLS92}\cite{BachmairGLS95}\cite{Weidenbach01}\cite{BachmairG01}\cite{NieuwenhuisR01}.

Compared to the existing approaches, we use a more expressive and implicit constraint language,
our search is guided by backjumping and learning non-redundant clauses, and 
our model representation is more compact, in general.
In addition, compared to all existing approaches, we can prove that all our learned clauses are non-redundant
and this way, for the first time, establish a calculus that combines the search with respect to a dynamically
changing (partial) model with an overall notion of redundancy. 
For a more detailed comparison, see Section 9.
%
%
%
%

In the rest of the paper, we first introduce some basic definitions and notions in Section 2, followed by 
a description of our calculus in Section 3.
Section 4 establishes its soundness, while, after introducing some regularity conditions in Section 5, we 
provide our key result, namely non-redundant clause learning, in Section 6. 
We then prove termination in Section 7.

In Section 8, we specify some details on handling constraints, and basic heuristics for a future implementation.
We compare our calculus to the existing literature in more details in Section 9.
Finally, Section 10 provides a summary and outlines future work.
%

%% file: src/s2-defs/definitions.tex
\section{Preliminaries}

\subsection{Basic Definitions}

We assume the reader is familiar with first-order logic, its syntax, and its semantics. 
In particular, we handle the {\mEPR} fragment, or {\mEPRs} for short
In this fragment the only functions allowed in the clause normal form are finitely many constants.
We denote the finite \emph{signature} by $\Sigma$, the \emph{set of predicate symbols} by $\operatorname{Pr}$, 
and call the finite set of constants the \emph{domain}, denoted by $\mDomain$.  

%

We denote the set of all first-order atoms over a signature $\Sigma$ and a possibly infinite set of variables $\text{X}$ by 
$\mathcal{A}_{\Sigma}(\text{X})$. 
In particular the set of ground atoms is denoted by $\mathcal{A}_{\Sigma}$, a short-hand for $\mathcal{A}_{\Sigma}(\emptyset)$.
For a literal $L$, $|L|$ denotes the atom contained by $L$.
In general, we denote the ground instances of an \emph{expression} - a term, literal, or clause - 
$e$ over the domain $\mDomain$ by the notation $\mGnd(e)$. 

W.l.o.g., we assume that each independent expression is variable disjoint, and we call a variable \emph{fresh} if it does not occur in any expression - e.g.\ clause or clause set - of the current context. 

We consider \emph{substitutions} in the usual way, and for a substitution $\sigma$, $\mDmn(\sigma)$ denotes the \emph{domain} of $\sigma$, i.e.\
the finite set of variables with $x \ne x\sigma$, and $\mRng(\sigma)$ denotes the \emph{range} of $\sigma$, i.e.\
the image of $\mDmn(\sigma)$ w.r.t.\ $\sigma$.

We assume the reader is familiar with \emph{most general unifiers}, 
and $\mMGU$ is used to denote the result of unifying two or more
expressions or substitutions.
We use the short-hand $\exists \sigma = \mMGU(e_1, e_2)$ to both state the existence of a most general unifier and bind 
$\sigma$ to one such substitution.


For expressions or substitutions $e_1$, $e_2$, we say $e_2$ \emph{can be matched against} $e_1$, or $e_1$ \emph{is more general than} $e_2$, and 
write $e_1 \geq e_2$, if and only if there is a substitution $\sigma$ such that $e_1 = e_2\sigma$. 

We represent a first-order interpretation $I$ with the set $\{ A \in \mathcal{A}_{\Sigma}~|~I \models A\}$.
We define \emph{satisfiability} and \emph{semantic consequence} as usual.

In particular, we consider the problem of deciding whether a finite clause set $\text{N}$ over a {\mEPRs} language $\Sigma$ without equality is satisfiable. 
This problem is known to be NEXPTIME-complete~\cite{Lewis80}.

\subsection{Constraints and Constrained Literals}
\input{src/s2-defs/ss2-constraints}
%
%

\subsection{Operations on Constrained Literals}
\input{src/s2-defs/ss3-operations}
%

\subsection{Model Representation}
\input{src/s2-defs/ss4-models}
%
%
%

\subsection{Induced Ordering}
\input{src/s2-defs/ss5-induced}
%
%

%% file: src/s2-defs/ss2-constraints.tex
%
Next, we provide details about the constraint language we use. Our constraints are equivalent with \emph{implicit generalizations}, a constraint language 
for representing terms and models with exceptions. 
It has applications in inductive learning, logic programming and term rewriting. 
For more details see e.g.\ 
\cite{Comon91}\cite{LassezM87}.

The name \emph{dismatching constraints} was chosen in the spirit of \emph{iProver}\cite{InstaGen13}, although for our purposes checking satisfiability 
has to be carried out over the ground instances and thus, the linear-time algorithm of \emph{iProver} based on matching is not applicable.

While implicit generalizations maintain a list of literals with fresh variables representing exceptions for the literal constrained, 
dismatching constraints extract the arguments of the literals and represent the restrictions as conjunctions of disequations to 
allow more simplification and a more compact representation.
In particular, we maintain a strict normal form, which already assumes most inexpensive simplifications.

We chose dismatching constraints for a balance between expressiveness and simplicity, for the existing literature, and for compactness. 
However, NRCL is compatible with any constraint language allowing the operations discussed in the next subsection.

%
\begin{defi}[Dismatching Constraint]\label{dmcDef}
A \emph{dismatching constraint} $\pi$ is of the form
\[\land_{i \in \mathcal{I}}~\vec{s}_i \ne \vec{t}_i\]
where $\mathcal{I}$ is a finite set of indices, and for each $i \in \mathcal{I}$, $\vec{s}_i$ and $\vec{t}_i$ are tuples of terms of the same dimension. 

Furthermore, we assume that all the left-hand side variables in $\pi$ differ from any right-hand side variable, 
and for each $i, j \in \mathcal{I}$, $\vec{t}_i$ and  $\vec{t}_j$ are variable disjoint whenever $i$ differs from $j$.

We further extend the set of constraints with the constants $\top$, $\bot$ representing the tautological and the unsatisfiable constraint, respectively.

Finally, an \emph{atomic constraint} $\vec{s} \ne \vec{t}$ occurring in $\pi$ is also called \emph{a subconstraint of $\pi$}.
\end{defi}
\noindent
To enforce a normal form, we make further assumptions below.
\begin{defi}[Normal Form]
We say a constraint $\pi = \land_{i \in \mathcal{I}}~\vec{s}_i \ne \vec{t}_i$ is \emph{in normal form} iff
the following conditions hold:
\begin{enumerate}[($C$1)]
	\item each $\vec{s}_i$ contains only variables
	\item no variable occurs more than once in any left-hand side $\vec{s}_i$
\end{enumerate}
\end{defi}

A simple consequence of the normal form is that the two sides of any subconstraint $\vec{s} \ne \vec{t}$ are 
always unifiable, and the \emph{induced substitution} $\{ \vec{s} \gets \vec{t} \}$ is always well-defined and 
matches the left-hand side against the right-hand side.

\begin{defi}[Induced Substitutions]
The \emph{set of induced substitutions} of a dismatching constraint $\pi$ in normal form is the set given by
\[\{ \{\vec{s}_i \gets \vec{t}_i\}~|~i \in \mathcal{I} \}\]
if $\pi = \land_{i \in \mathcal{I}}~\vec{s}_i \ne \vec{t}_i$. For $\bot$, we define 
it as the set containing only the identity, and for $\top$ as the empty set.
\end{defi}

We define $\mLVar(\pi)$ and $\mRVar(\pi)$ as the set of the left-hand side and right-hand side variables of some dismatching constraint $\pi$, respectively. 
Then the semantics for our constraints can be given as below.

\begin{defi}
A \emph{solution} of a constraint $\pi$ over some variable set $V$, which contains $\mLVar(\pi)$ but contains no variable from $\mRVar(\pi)$, 
is a ground substitution $\delta: V \rightarrow \mDomain$ such that 
no $\vec{t}_i$ can be matched against the respective $\vec{s}_i\delta$, i.e.\
no $\vec{s}_i\delta$ is an instance of the respective $\vec{t}_i$.

In particular, if $\pi = \top$, any such grounding substitution is a solution, and $\pi = \bot$ has no solution at all.
\end{defi}

\noindent
As usual, $\pi$ is called \emph{satisfiable} and \emph{unsatisfiable} if it has a solution or no solution, respectively. 
We note that the notion of satisfiability depends only on $\mLVar(\pi)$.

\begin{example} Consider the domain $\mDomain = \{a, b\}$ and the constraint 
$$\pi = (x,y) \ne (v,v) \land y \ne a$$
Then $\pi$ is satisfiable and the ground substitution $\sigma = \{x \gets a, y \gets b\}$ 
is the only solution of $\pi$ (over $V = \{x,y\}$), since $y$ can only be $b$ and the first 
subconstraint represents $x \ne y$.
\end{example}

\begin{remark}\label{a-inducedsubst-remark}
It can be shown that a ground substitution $\delta: V \rightarrow \mDomain$ with $\mLVar(\pi) \subseteq V$ 
is not a solution of $\pi$ 
if and only if there is an induced substitution $\sigma$ which is more general than $\delta$.
\end{remark}

\begin{defi}
Let $\pi$ and $\pi'$ denote constraints for which both
\begin{itemize}
\item $\mLVar(\pi)\cap\mRVar(\pi') = \emptyset$, and
\item $\mLVar(\pi')\cap\mRVar(\pi) = \emptyset$ 
\end{itemize}
hold.
Such constraints are called \emph{equivalent} iff their sets of solutions coincide for any $V$ such that 
$\mLVar(\pi) \cup \mLVar(\pi') \subseteq V$, and both
$V \cap \mRVar(\pi) = \emptyset$ and 
$V \cap \mRVar(\pi') = \emptyset$.
\end{defi}

\subsubsection*{Normal Form Transformation}
Next, we show that any dismatching constraint of the form $\land_{i \in \mathcal{I}}~\vec{s}_i \ne \vec{t}_i$ 
can be normalized in polynomial time. 
This can be achieved with the rule set below, given as rewriting rules over the subconstraints.

\begin{enumerate}
	\item $(\vec{s}_1, a, \vec{s}_2) \ne (\vec{t}_1, a, \vec{t}_2) \Rightarrow (\vec{s}_1, \vec{s}_2) \ne (\vec{t}_1, \vec{t}_2)$, 
		where $a \in \mDomain$
	\item $(\vec{s}_1, a, \vec{s}_2) \ne (\vec{t}_1, b, \vec{t}_2) \Rightarrow \top$, 
		where $a \ne b \in \mDomain$
	\item $(\vec{s}_1, a, \vec{s}_2) \ne (\vec{t}_1, x, \vec{t}_2) \Rightarrow (\vec{s}_1, \vec{s}_2) \ne (\vec{t}_1, \vec{t}_2)\sigma$, 
		where $a \in \mDomain$, $\sigma = \{ x \gets a\}$
	\item $() \ne () \Rightarrow \bot$
	\item $(\vec{s}_1, x, \vec{s}_2, x, \vec{s}_3) \ne (\vec{t}_1, r_1, \vec{t}_2, r_2, \vec{t}_3) \Rightarrow (\vec{s}_1, x, \vec{s}_2, \vec{s}_3) \ne (\vec{t}_1, r_1, \vec{t}_2, \vec{t}_3)\sigma$, 
		if $\exists \sigma = \mMGU(r_1,r_2)$
	\item $(\vec{s}_1, x, \vec{s}_2, x, \vec{s}_3) \ne (\vec{t}_1, r_1, \vec{t}_2, r_2, \vec{t}_3) \Rightarrow \top$, 
		if $\nexists\mMGU(r_1,r_2)$
	\item $\vec{s} \ne \vec{t} \Rightarrow \bot$, 
		if $\vec{t}$ can be matched against $\vec{s}$
	\item $(\vec{s}_1, \vec{s}_2) \ne (\vec{t}_1, \vec{t}_2) \Rightarrow \vec{s}_1 \ne \vec{t}_1$,
		if $\mVar(\vec{t}_1) \cap \mVar(\vec{t}_2) = \emptyset$, and $\vec{t_2}$ can be matched against $\vec{s}_2$ 
		
\end{enumerate}
Where the last rule is considered modulo permutations of positions corresponding to the $(\vec{s}_1, \vec{s}_2)$-partitionings.
\begin{example} Let us normalize the following constraint: 
\[(x,a,y,x) \ne (b, v, w, w) \land (x,a,y,x) \ne (w_0, w_0, v_0, t_0)\]
For the first subconstraint we get
\[(x,a,y,x) \ne (b, v, w, w) \stackrel{(3)}{\Rightarrow} (x,y,x) \ne (b, w, w) \stackrel{(4)}{\Rightarrow} (x,y) \ne (b,b)\]
and for the second one
\[(x,a,y,x) \ne (w_0, w_0, v_0, t_0) \stackrel{(4)}{\Rightarrow} (x,a,y) \ne (w_0, w_0, v_0) 
\stackrel{(3)}{\Rightarrow} (x,y) \ne (a, v_0) \stackrel{(8)}{\Rightarrow} x \ne a\]
Thus, the normalized constraint is
\[(x,y) \ne (b,b) \land x \ne a\]
\end{example}
Applying these rules together with the usual rules for conjunction and the constants $\bot, \top$
\begin{enumerate}
	\item preserves the variable disjointness conditions of Definition~\ref{dmcDef}
	\item preserves solutions, i.e.\ the left-hand side and right-hand side constraints are equivalent
	\item transforms $\pi$ into normal form in polynomial time 
\end{enumerate}
We note that the rules (7) and (8) are optional, and that (7) is a special case of (8).
%

Therefore, w.l.o.g. we assume that the constraints are always in normal form, and 
the result of any operation is transformed into normal form without explicitly expressing it. 
We also express it by using the notation $\land_{i \in \mathcal{I}}~\vec{x}_i \ne \vec{t}_i$ 
for dismatching constraints in the rest of the paper.
%
\subsubsection*{Constrained Literals}
Next, we define literals constrained with dismatching constraints in normal form, and give their semantics as sets of 
ground literals.

\begin{defi}[Constrained Literal]
We call the pair $(L; \pi)$ of a literal $L$ and a dismatching constraint $\pi$ such that both
$\mLVar(\pi) \subseteq \mVar(L)$ and 
$\mRVar(\pi) \cap \mVar(L) = \emptyset$ hold
a \emph{constrained literal}.

The semantics of constrained literals is given by the following definition of the \emph{set of covered literals}:
\[\mGnd(L; \pi) = \{L\delta~|~\delta: \mVar(L)\rightarrow\mDomain\text{~s.t.~} 
					\forall i \in \mathcal{I}: \nexists\mMGU(\vec{x}_i\delta, \vec{t}_i)\}\]
where $\pi = \land_{i \in \mathcal{I}} \vec{x}_i \ne \vec{t}_i$.
A ground literal $L'$ is \emph{covered by} a constrained literal $(L; \pi)$ iff $L' \in \mGnd(L;\pi)$.

We say that a constrained literal $(L; \pi)$ is \emph{empty} if it covers no ground literals, i.e.\
$\mGnd(L; \pi)$ is empty.
\end{defi}

\indent
It is easy to see that $(L; \pi)$ is empty if and only if $\pi$ is unsatisfiable, and that given a solution $\delta$ of $\pi$ over 
$\mLVar(\pi)$, for any extension $\delta'$ of $\delta$ to $\mVar(L)$, 
$L\delta' \in \mGnd(L; \pi)$ holds.

\begin{example} Let $(L; \pi) = (P(x,y);~(x,y) \ne (v,v) \land x \ne a \land y \ne b)$.
Then the set of covered literals over the domain $\mDomain_2 = \{ a, b\}$ is
\[\mGnd(L; \pi) = \{ P(b,a) \}\]
and if we take $\mathcal{D}_3 = \{a,b,c\}$ instead, it is
\[\mGnd(L; \pi) = \{ P(b,a), P(c,a), P(b,c) \}\]
\end{example}
%
\noindent
In the rest of the paper we make some further assumptions as common in automated reasoning:
\begin{enumerate} 
	\item Different constrained literals are variable disjoint, unless stated otherwise.
	\item Apart from normal form transformations, for any substitution $\sigma$ applied to a constrained literal $(L; \pi)$,
	the following always hold unless stated otherwise:
	\begin{itemize}
		\item $\mDmn(\sigma) \cap \mRVar(\pi) = \emptyset$
		\item $\mVar(\mRng(\sigma)) \cap \mRVar(\pi) = \emptyset$
	\end{itemize}
\end{enumerate}
\subsubsection*{Constrained Clauses}
Occasionally, we have to represent a collection of ground clauses by a constrained clause $(C; \pi)$. 
Extending the notations and semantics for constrained literals to constrained clauses is straightforward. 

Furthermore, we use the notation $(C; \sigma; \pi)$ for the constrained clause $(C\sigma; \pi)$, whenever we 
wish to syntactically distinguish $C$ and $\sigma$.

We only note that during resolving away literals from $C$, we might get to a state where 
$\mLVar(\pi)$ contains variables not occurring in $C$. See the constrained unit clause
\[(P(y,z); (x,y) \ne (v,v) \land (x,z) \ne (w,w))\]
from Example~\ref{freeVarEx} for a demonstration.

For semantic purposes, these \emph{free variables} are considered existential variables.
We assume that such variables are eliminated through instantiation, see Section 8 for further details.

%% file: src/s2-defs/ss3-operations.tex
%
In the context of our calculus, three operations are of significance: conjunction, difference, 
and checking whether a constrained literal is empty. 

%
%

In the literature checking emptiness also relates to \emph{sufficient completeness} and 
\emph{negation elimination}
and it is known to be a co-NP-complete problem~\cite{LassezM87} in the case of finitely many function symbols and infinite Herbrand universe.

This complexity result also holds for our setting - 
one might take a binary domain with $true$ and $false$, and then each atomic constraint with constant right-hand side 
can be seen as clauses with the left-hand side variables as propositional variables, 
and the emptiness of the whole constraint as the unsatisfiability of this clause set.

In this section, we propose an enumeration-based algorithm to test emptiness as an alternative
to relying on external CSP and CDCL solvers.

\subsubsection*{Conjunction}

For two constrained literals $(L_1; \pi_1)$, $(L_2; \pi_2)$ with the same polarity and predicate symbol, we look for a constrained literal $(L; \pi)$ for which
\[\mGnd(L; \pi) = \mGnd(L_1; \pi_1) \cap \mGnd(L_2; \pi_2)\]
holds. If the two literals are unifiable,   
such a literal exists. Otherwise, any empty constrained literal can be chosen.

\begin{defi}[Conjunction] 
Let as the define and denote \emph{the conjunction of two constrained literals $(L_1; \pi_1)$, $(L_2; \pi_2)$} as
\[(L_1; \pi_1) \land (L_2; \pi_2) = (L_1\sigma; \pi_1\sigma \land \pi_2\sigma)\]
if $\exists \sigma = \mMGU(L_1, L_2)$. If the literals are not unifiable, we define it as the empty $(L_1; \bot)$.
\end{defi}

\noindent
This definition is sound, i.e.\

\begin{lemm} For any unifiable constrained literals $(L_1; \pi_1)$, $(L_2; \pi_2)$,
\[\mGnd(L_1; \pi_1) \cap \mGnd(L_2; \pi_2) = \mGnd(L_1\sigma; \pi_1\sigma \land \pi_2\sigma)\]
holds, where $\sigma = \mMGU(L_1, L_2)$.
\end{lemm}

\noindent
\startproof 

($\subseteq$): Consider a ground literal from $\mGnd(L_1; \pi_1) \cap \mGnd(L_2; \pi_2)$, and w.l.o.g. assume it has the form $L_1\delta$. 
Then $L_1\delta \geq L_i$ holds for both $i= 1,2$.

Thus, $\delta = \sigma\epsilon$ for some substitution $\epsilon$.  Since $L_1\delta \in \mGnd(L_i; \pi_i)$, $\pi_i\sigma\epsilon$ must be true ($i = 1,2$).
But then $L_1\delta \geq L_1\sigma$, and $L_1\delta = (L_1\sigma)\epsilon \in \mGnd(L_1\sigma; \pi_1\sigma \land \pi_2\sigma)$ both hold.

($\supseteq$): Now, assume that $(L_1\sigma)\epsilon$ is a literal from $\mGnd(L_1\sigma; \pi_1\sigma \land \pi_2\sigma)$.
Then, since $\sigma$ is the most general unifier, we know that $L_1\sigma\epsilon \geq L_i$ hold for both $i = 1,2$.
Furthermore, $\pi_i\sigma\epsilon$ is true ($i = 1,2$), and thus, $L_1\sigma\epsilon \in \mGnd(L_1; \pi_1)$ and 
$L_1\sigma\epsilon = L_2\sigma\epsilon \in \mGnd(L_2; \pi_2)$.
\finishproof\leaveabit\noindent
We note that the case when no unifier exists is trivial.

\begin{example} Consider the following constrained literals
	\begin{itemize}
		\item $(L; \pi) = (P(x,y);~(x,y) \ne (v,v) \land x \ne a \land y \ne b)$
		\item $(L'; \pi') = (P(z,a);~z \ne b)$
	\end{itemize}
Then according to the definition above
\[(L; \pi) \land (L'; \pi') = (P(z,a);~(z,a) \ne (v,v) \land z \ne a \land a \ne b \land z \ne b)\]
which can be simplified to
\[(P(z,a); z \ne a \land z \ne b)\]
This expression is empty over $\mDomain_2 = \{a,b\}$, and covers exactly the atom $P(c,a)$ over $\mDomain_3 = \{ a,b,c \}$.
\end{example}

\subsubsection*{Difference}

The  \emph{difference}, or \emph{relative difference}, $(L; \pi)$ of two constrained literals $(L_1; \pi_1)$, $(L_2; \pi_2)$ satisfies 
\[\mGnd(L; \pi) = \mGnd(L_1; \pi_1) - \mGnd(L_2; \pi_2)\]

Again, if the two literals are unifiable, such a $\pi$ does exist for any finite domain - in the worst case we just add ground constraints to rule out the disallowed atoms.
However, this operation might increase the size of $\pi$ exponentially, as demonstrated by the example below.

\begin{example}\label{a-dmcDiffEx1}
Consider the difference 
\[(L(x_1, x_2, x_3); \top) - (L(x_1, x_2, x_3); \land_{i=1}^3~x_i \ne a)\]
where $\operatorname{arity}(L) = 3$. 
If $\mDomain_2 = \{ a, b \}$, we might get the still simple expression 
\[(L(x_1, x_2, x_3); (x_1, x_2, x_3) \ne (b,b,b))\]
However, if $\mDomain_3 = \{ a, b, c \}$, the best we can get is 
\[(L(x_1, x_2, x_3); (x_1,x_2,x_3) \ne (b,b,b) \land (x_1,x_2,x_3) \ne (c, b, b) \land \dots \ne (c,c,c))\]
It is easy to see that in general, if $|\mDomain| = n$ with $a \in \mDomain$, and $\operatorname{arity}(L) = r$, 
the size of the resulting constraint is $O((n-1)^r)$.
\end{example}

Alternatively, one might take a set of disjoint constrained literals describing the difference as follows. 
First, take the simpler case when $L_1$ and  $L_2$ are the same literal $L$, and consider the difference $(L; \pi_1) - (L; \pi_2)$. 
Assume $\pi_1 = \land_{i \in \mathcal{I}_1} \nu_i$, $\pi_2 = \land_{i \in \mathcal{I}_2} \eta_i$, 
and $\{\sigma_i~|~i \in \mathcal{I}_2\}$ is the set of induced substitutions for $\pi_2$.
Then, the constrained literal set 
\[\{ (L\sigma_i; \pi_1\sigma_i)~|~i \in \mathcal{I}_2 \}\]
describes the difference, i.e.\

\begin{lemm}
\[\bigcup_{i \in \mathcal{I}_2}\mGnd(L\sigma_i; \pi_1\sigma_i) = \mGnd(L; \pi_1) - \mGnd(L; \pi_2)\]
\end{lemm}

\noindent
\startproof 

($\supseteq$): Assume $L\delta \in (\mGnd(L; \pi_1) - \mGnd(L; \pi_2))$. 
Since $L\delta\notin\mGnd(L; \pi_2)$, a subconstraint $\eta_i \in \pi_2$ must be violated, i.e.\ 
for some $i \in \mathcal{I}_2$, $\eta_i\delta = \bot$. 

Then, by the earlier Remark~\ref{a-inducedsubst-remark}, $\delta \geq  \sigma_i$ where $\sigma_i$ is the corresponding induced substitution. Thus,
$\delta = \sigma_i\xi$ for some substitution $\xi$. 
Finally, since $\pi_1\delta = (\pi_1\sigma_i)\xi$, $L\delta \in \mGnd(L\sigma_i; \pi_1\sigma_i)$ must hold.

($\subseteq$): Now, assume $(L\sigma_i)\xi \in \mGnd(L\sigma_i; \pi_1\sigma_i)$ 
for some $i \in \mathcal{I}_2$ and grounding substitution $\xi$.
Then, we know that $\pi_1\sigma_i\xi = \top$, and that $\pi_2\sigma_i\xi = \bot$ since $\eta_i\sigma_i\xi = \bot$.
Thus, $L\sigma_i\xi \in (\mGnd(L; \pi_1) - \mGnd(L; \pi_2))$.
\finishproof\\[6 pt]
\noindent
However, this set is not pairwise disjoint, and therefore a further step is needed for our purposes.

\begin{lemm}\label{diffLemma2}
W.l.o.g. assume $\mathcal{I}_2 = \{1, \dots, l\}$, and take
\[\{ (L\sigma_i; \pi_1\sigma_i \land \eta_1\sigma_i \land \eta_2\sigma_i \land \dots \land \eta_{i-1}\sigma_i)~|~i = 1, \dots, l\}\]
Then this set still describes the difference and its elements are pairwise disjoint.
\end{lemm}

\noindent
\startproof We only prove one inclusion, as the other direction is analogous to the first proof, and 
disjointness trivially follows form the definition of the set.

($\supseteq$): Assume $L\delta$ is a ground literal from the difference. Thus, $\pi_1\delta = \top$ and 
$\eta_i\delta = \bot$ for at least one $i \in \mathcal{I}_2$. Let $i$ be the smallest (left-most) such index.

Then, $\delta \geq \sigma_i$ must hold along with $\eta_j\delta = \top$ for each $j < i$ from $\mathcal{I}_2$. Thus,
$L\delta \in \mGnd(L\sigma_i; \pi_1\sigma_i \land \eta_1\sigma_i \land \eta_2\sigma_i \land \dots \land \eta_{i-1}\sigma_i)$.
\finishproof\\[6 pt]
\indent
We also note that the above manipulations preserve the variable disjointness of the left-hand and right-hand sides.

\begin{example}
Carrying on with example~\ref{a-dmcDiffEx1} above, we have $\mathcal{I}_2 = \{ 1, 2, 3\}$, $\eta_i: x_i \ne a$ and $\sigma_i = \{ x_i \gets a\}$, which gives
\[\{ (L(a, x_2, x_3); \top), (L(x_1,a, x_3); x_1 \ne a), (L(x_1, x_2, a); x_1 \ne a \land x_2 \ne a)\}\]
as a result.
\end{example}

\indent
Let $|\pi|$ denote the size of $\pi$. 
Then, this operation introduces $O(|\mathcal{I}_2|)$ atoms with a maximal constraint size of $O(|\pi_1| + |\pi_2|)$ in general. 
This gives a total size of $O(2|\pi^*|^2)$ where $|\pi^*| = \operatorname{max}\{|\pi_1|, |\pi_2|\}$.
Clearly, it is independent of the domain size.

\begin{lemm}
Finally, if $L_1 \ne L_2$, but $\mMGU(L_1, L_2) = \delta$ exists (otherwise the difference is $(L_1; \pi_1)$), and $\vec{u}$ denotes the argument of the top symbol in $L_1$, we get the desired set by adding 
\[(L_1; \pi_1 \land \vec{u} \ne \vec{u}\delta\rho)\]
to the set $(L_1\delta; \pi_1\delta) - (L_2\delta; \pi_2\delta)$ 
where the variable renaming $\rho$ introduces fresh variables for 
the variables in $\vec{u}\delta$.
\end{lemm}

\startproof The literals in $\mGnd(L_1; \pi_1)$ can be divided into two disjoint group based on whether they are instances of $L_2$ or not.

Those that are no instances of $L_2$ are covered by the proposed constrained literal $(L_1; \pi_1 \land \vec{u} \ne \vec{u}\delta\rho)$. Clearly, 
each such literal is in the difference.

The common instances are covered by $(L_1\delta; \pi_1\delta)$. From these literals we have to remove those which are covered by $(L_2; \pi_2)$ as well. 
Clearly, it is enough to compute the difference $(L_1\delta; \pi_1\delta) - (L_2\delta; \pi_2\delta)$.

The resulting set together with $(L_1; \pi_1 \land \vec{u} \ne \vec{u}\delta\rho)$ covers exactly the elements of the difference.
\finishproof\\[6 pt]
\indent
It is easy to see that the proofs above hold even if some of the constraints are the constants $\bot$ or $\top$, 
and our definition of induced substitutions in the case of constants supports the proofs.

The disadvantage of the second method is the fragmentation of the constrained literal, 
especially since after every time we derive a new assignment we have to subtract each unifiable already defined literal.

\begin{remark}
Whenever we compute a difference and get a set of literals as a result, we carry on working with the literals separately. 

We could extend our constraint language to handle a set of constrained literals as a single expression. 
In the literature the corresponding constraints are called \emph{disjunctive implicit generalizations}, see e.g.\
\cite{Pichler03} for details.
\end{remark}

\subsubsection*{Checking Emptiness}

The problem of deciding whether a constrained literal is empty. As we mentioned before, this is equivalent with the unsatisfiability of the corresponding constraint. 

This is in general a co-NP-complete problem~\cite{Comon91}\cite{LassezM87}. 
Lassez and Marriott proposes an algorithm for computing explicit representation in~\cite{LassezM87}, which can be used for determining emptiness as well. 
Their algorithm is based on generating disjoint partitions of instances by instantiating a single variable with every possible function symbol at every step.

We note that the operation is indeed complex, 
but so is checking subsumption and subsumption resolution in first-order theorem provers, 
and even \emph{iProver} calls CDCL iteratively. 
Yet, these techniques are efficient in practice, which we consider an indication that 
an efficient implementation of NRCL is possible. 

%
We propose here an enumeration-based algorithm. 
Assume $\mDomain = \{a_1, \dots, a_n\}$ is ordered by $>$,
 and $a_n > a_{n-1} > \dots > a_0$.
For a constrained literal $(L; \pi)$ with left-hand side variables $x_1, \dots, x_k$, we 
find a solution $(c_1, c_2, \dots, c_k)$ denoting $\{ x_j \gets c_j~|~j = 1, \dots, k\}$ by
enumerating the possible assignments starting with $(a_0, a_0, \dots, a_0)$.

If for an intermediate assignment $(c_1, c_2, \dots, c_k)$ the subconstraint $\vec{x} \ne \vec{t} \in \pi$
is false, then we increase the \emph{value} of the right-most position involved in $\vec{x}$.
If it is already $a_n$, we reset it to $a_0$ and increase the next involved variable to the left. 

If no further increase is possible, there is no solution. 
If we get a solution for $\vec{x} \ne \vec{t}$, we pick the 
left-most involved variable which we changed, and reset all non-$\vec{x}$ variables to $a_0$.

By repeating the above steps, we either get a solution satisfying $\pi$, or attempt to increase beyond $(a_n, a_n, \dots, a_n)$, proving the 
unsatisfiability of the constraint.

Besides simplicity, this algorithm also has the advantage that the solution might be reusable in operations. 
We only need to make sure that the solutions for the operands are comparable in the sense that 
they denote the minimal solutions of the respective constraints w.r.t.\
the same ordering over the possible assignments.

This can be ensured by ordering the variables based on their left-most occurrence. 
This way, the solutions of any two non-empty constrained literals with the same predicate symbol are comparable by taking 
the arguments of the ground literals representing the solutions.

Then, it is enough to consider substitution and adding new subconstraints as primitive operations. Both meet and difference builds upon these steps. 

When applying a substitution $\sigma$, we check if the current solution satisfies the positive equality constraints induced by $\sigma$. 
If yes, then we apply the substitution and keep this solution.

Otherwise, we keep enumerating by always checking the positive conditions first. If we find an assignment satisfying both $\pi$ and $\sigma$, we apply $\sigma$ and 
save the new solutions. If no solutions to be found, the new constrained literal is empty.

When extending $\pi$ with a new subconstraint $\vec{x} \ne \vec{t}$, 
we simply continue the enumeration with the current solution and the extended constraint $\pi \land \vec{x} \ne \vec{t}$.

%% file: src/s2-defs/ss4-models.tex
%
\subsubsection*{Model Candidate}
On the course of this paper, we represent a \emph{model candidate}, also called a \emph{model assumption}, as a set $\Gamma$ of constrained literals. 

\begin{defi}
A set of constrained literals $\Gamma$ is called \emph{consistent} if there is no ground atom covered by both a positive and a negative literal from $\Gamma$.

$\Gamma$ is \emph{strongly consistent} if its elements are pairwise disjoint w.r.t.\
covering atoms, i.e.\ 
for all different $(L;\pi), (L'; \pi') \in \Gamma$,
$\mGnd(|L|; \pi) \cap \mGnd(|L'|; \pi') = \emptyset$.
\end{defi}

\noindent
We consider only strongly consistent sets in this paper.

\begin{defi}[Induced Interpretation]
The set of positive constrained literals in $\Gamma$ is denoted by $\Gamma^+$. 
Then the \emph{first-order interpretation $I_{\Gamma}$ induced by} $\Gamma$ is given as
\[I_{\Gamma} = \bigcup_{(L;\pi) \in \Gamma^+}~\mGnd(L;\pi)\]
\end{defi}

This interpretation serves as a minimal model defined by the positive literals, and it is
used in the rule \emph{Success} and the relevant proofs.

\subsubsection*{Trail}

NRCL attempts to lift the classic CDCL, and as such, it uses a sequence of literals to store the current partial model assumption. 

This \emph{trail} in our case is a sequence of annotated constrained literals. 
We retain the notation $\Gamma$, and extend all our definitions and operations for sets of constrained literals to trails as well. 
We call the elements of $\Gamma$ \emph{assignments}, as they define truth-values of ground atoms.

Literals in $\Gamma$ are either \emph{decision} or \emph{deduced literals}. Decisions are annotated with a unique positive integer, with $(L; \pi)^i$
representing the $i$th decision in $\Gamma$. 
Deduced literals are annotated with their reasons, a first-order clause from the current clause set.
In the course of the paper, $\alpha$ is used to denote an arbitrary annotation, $C$ to denote a reason clause, and $k, l, i$ to denote integers.

We define the value of a ground literal or ground clause \emph{true}, \emph{false}, or \emph{undefined} under $\Gamma$
lifting the notions of CDCL. 
In particular, a ground literal $L'$ \emph{is defined by} a constrained literal $(L; \pi) \in \Gamma$, iff
$|L'| \in \mGnd(|L|;\pi)$. If such an $(L; \pi)$ exists, we also say that $\Gamma$ \emph{defines} $L'$.
Then, the value of the defined ground literal $L'$ is \emph{true} iff $L'$ and $L$ has the same polarity.

Non-ground literals are treated as unit clauses, and a set of ground clauses represented 
by the constrained clause $(C; \pi)$ is \emph{true} or \emph{false} in $\Gamma$, 
if all of the covered ground 
instances are \emph{true}, or \emph{false}, respectively.
The notion of \emph{defined by} $\Gamma$ extends to constrained clauses similarly, \ie
$(C; \sigma)$ is defined \wrt $\Gamma$ iff for each $C' \in \mGnd(C; \sigma)$, at least 
one $L' \in C'$ is defined under $\Gamma$.
We note that the definition of a \emph{false} constrained clause is non-standard, and 
it is formulated this way to conveniently define later the invariants and the rule \emph{Conflict}.
The notion of \emph{defined by} $\Gamma$ can be extended to constrained clauses similarly.

\emph{The level $\operatorname{lvl}(L)$ of a ground literal $L$ w.r.t.\
the trail $\Gamma$} is defined as in CDCL: 
The annotation of a decision in $\Gamma$ is the \emph{level} of this decision literal.
Then, the \emph{level $\operatorname{lvl}(L)$ of a defined ground literal $L$ w.r.t.\
$\Gamma$} is the level of the last decision in $\Gamma$ before the constrained literal defining $L$, and 
zero if no such decision exists.

If $k$ is the level of a literal, we might also say \emph{the literal is of level $k$}.
We call the largest level occurring in a trail the \emph{top-level}, and also the level of the trail. 
If no decision occurs in the trail, it is considered $0$.

Following the terminology of SAT solvers, we call a ground clause \emph{assertive} iff it is false w.r.t.\
the current trail and contains exactly one top-level literal.

Finally, we say a first-order clause $C$ or a constrained clause $(C\sigma; \pi)$ \emph{is assertive} iff $\mGnd(C)$, and $\mGnd(C\sigma; \pi)$ contains 
at least one assertive ground clause, respectively.

\subsubsection*{Induced Abstraction}
Using $\Gamma$ to define truth-values for groups of ground atoms represented by constrained literals can also be seen as providing a propositional 
abstraction and an abstract partial interpretation. 

In this context, our calculus can be seen as a fine-grained abstraction-refinement algorithm, which interleaves refinement and abstract model search, 
and lets the clauses and decision heuristics guide the implicit abstraction and refinement steps. 

Below, we provide the related definitions and use these later to define our induced ordering.
Beyond this, we do not take any advantage of this connection. Further investigation this direction and 
utilizing existing results for abstraction-refinement-based procedures is left for future work.

We call a set of positive constrained literals $\Phi$ an \emph{abstraction}. 
An abstraction $\Phi$ provides a (partial) partitioning of $\mathcal{A}_{\Sigma}$, 
and by identifying its elements with propositional atoms, we can assign a propositional abstraction to our clause set $\text{N}$.

These propositional atoms are called \emph{abstract atoms}. 
The notions \emph{abstract literal} and \emph{abstract clause} are the corresponding syntactic expressions built from abstract atoms.
We use the \emph{abstraction function} $\mDef_{\Phi}$, or simply $\mDef$, to assign the set of abstract expressions to literals or clauses w.r.t.\
an abstraction $\Phi$.

Then an \emph{abstract interpretation} over an abstraction $\Phi$ is a propositional interpretation over the corresponding abstract atoms. 

If the totality of $\mDef_{\Phi}$ is needed, we identify uncovered ground atoms with the unique abstract atom $\bot$, and the domain of the interpretation is 
extended accordingly.

The abstraction $\Phi_{\Gamma}$ induced by $\Gamma$ is defined as
\[\Phi_{\Gamma} = |\Gamma| = \{(|L|; \pi)~|~(L; \pi) \in \Gamma\}\]
If $\Gamma$ is strongly consistent, $\Phi_{\Gamma}$ is always consistent, i.e.\ 
any ground atom is covered by at most one element of $\Phi$.

$\Gamma$ can be seen as defining an abstract interpretation over $\Phi_{\Gamma}$ assigning truth-values to abstract 
atoms based on the polarity of the corresponding constrained literals in $\Gamma$, and $undef$ to the abstract atom $\bot$.
%

%% file: src/s2-defs/ss5-induced.tex
%
In the following, let $<$ denote a given well-founded total ordering over ground expressions - atoms, literals and clauses. 
Furthermore, let $\Gamma$ denote a strongly consistent trail.

\begin{defi} The \emph{abstraction function $\mDef$ defined by $\Gamma$} is given as
\[\mDef(P) = 
\left\{
	\begin{array}{ll}
		(L;\pi)  & \mbox{if } (L;\pi) \in \Gamma\text{~and~$(L;\pi)$ defines $P$} \\
		\bot & \mbox{if no such $(L;\pi) \in \Gamma$ exists }
	\end{array}
\right.
\]
for each $P \in \mathcal{A}_{\Sigma}$.\leaveabit
Then, $\mDef$ can be extended to ground literals and clauses by
assigning the corresponding negated abstract atom to a negative literal, and the disjunction 
of the corresponding abstract literals to a clause, respectively.
\end{defi}

\begin{defi} \emph{The precedence ordering $<_p^{\Gamma}$ ($<_p$) defined by $\Gamma$}
is the ordering over the constrained literals in $\Gamma$ defined by their position in $\Gamma$, i.e.\

$(L_1; \pi_1) <_p (L_2; \pi_2)$ iff 
\[\Gamma = \Gamma_1, (L_1; \pi_1)^{\alpha_1}, \Gamma_2, (L_2; \pi_2)^{\alpha_2}, \Gamma_3\]
for some $\Gamma_1, \Gamma_2, \Gamma_3$ and annotations $\alpha_1$, $\alpha_2$.

We extend the ordering to $\Gamma\cup\{\bot\}$ with $\bot$ as maximal element. 
Finally, this ordering is extended to abstract literals and clauses as usual.
\end{defi}


\begin{defi}\label{a-inducedOrderDef2}
\emph{The ordering $<^{atom}_{\Gamma}$ induced by $\Gamma$} is defined over $\mathcal{A}_{\Sigma}$ and given as follows: $P <^{atom}_{\Gamma} Q$ iff either
\begin{enumerate}
	\item $\mDef(P) <_p \mDef(Q)$, or
	\item $\mDef(P) = \mDef(Q)$ and $P < Q$
\end{enumerate}
The ordering is extended to ground literals in the usual way, resulting in the literal ordering $<^{lit}_{\Gamma}$. 
\leaveabit
Finally, we extend it to ground clauses: $C <_{\Gamma} C'$ iff either
\begin{enumerate}
	\item $\mDef(C) <_p \mDef(C')$, or
	\item $\mDef(C) = \mDef(C')$ and $C~(<^{lit}_{\Gamma})_{mul}~C'$
\end{enumerate}
where $(<^{lit}_{\Gamma})_{mul}$ denotes the multiset extension of the literal ordering. 

$<_{\Gamma}$ extends the atom and literal orderings, and 
we call it \emph{the ordering induced by $\Gamma$}.
\end{defi}

\begin{prop}
$<_{\Gamma}$ is well-defined, total on ground clauses, and a well-founded ordering.
\end{prop}
\startproof 
It is easy to see that both $<_p$ and $(<_{\Gamma}^{lit})_{mul}$ are well-founded and total orderings 
over ground clauses. 
Since $<_{\Gamma}$ is the lexicographical combination of these orderings, $<_{\Gamma}$ inherits these properties.
\finishproof\leaveabit\indent
This dynamic ordering captures the local correlation between 
the atoms and literals in the search, and shifts the focus on the 
recent behavior of the calculus.
Finally, we introduce an easily provable proposition, which is used in the proofs later on.

\begin{prop}\label{orderPropProp}
Let $<$ be \emph{an arbitrary well-founded and total ordering over ground clauses}, 
$S$ and $S'$ finite sets of ground clauses, and assume there is a function
$\gamma: S' \rightarrow S$ such that for each $C \in S'$, $C < \gamma(C)$.
\leaveabit
Then $S' < S$ holds w.r.t.\
the multiset extension of $<$.
\end{prop}
%

%% file: src/s3-calc/calculus.tex
\section{Calculus}

The calculus NRCL attempts to find a model through a series of both arbitrary and deduced assignments. 
Analogous to the propositional SAT solvers, we apply propagation to find literals implied by 
existing assignments, and once it is exhausted, we add arbitrary literals, so-called \emph{decisions} to the trail.

We call this phase \emph{conflict search} and it ends with either a model of the original clause set, or with finding 
a clause $C$ with some instances given in the form $(C; \sigma; \pi)$ falsified by the current trail.
In the latter case, we start \emph{conflict resolution} and through resolving the current false clause with reason clauses 
from the trail, we learn a new assertive clause and backtrack to a state where this clause is not yet falsified by removing some 
of the most recent assignments from the trail.

As opposed to propositional SAT solving, where every clause can be considered already exhaustively factorized, 
in our case some ground instances might be still subject to factorization, and this requires further rules.

The rule \emph{Factorize} handles this during clause learning.
However, the calculus might still reach a state where the right-most literal on the trail is the last decision, the learnable clause is not 
assertive, but no factorization is possible.
When such a state is reached, we simply learn the current candidate for clause learning.
To avoid this situation again, we further demand that a new decision should not falsify any clause instance immediately, unless
\emph{Factorize} is applicable.

We call a clause \emph{blocking} a new decision if adding the decision to the trail would falsify an instance of the clause without
allowing \emph{Factorize} to handle the immediate conflict, see the precise definition below. 
We note that a clause learned in the above fashion blocks the last decision.

\begin{defi}
We say that \emph{a decision $(L; \pi)$ is blocked in $\Gamma$ by a clause $C$}, if $C$ has 
a ground instance $C\sigma$ with $L_1, L_2 \in C\sigma$ such that for $\Gamma' = \Gamma, (L; \pi)$
\begin{itemize}
	\item $C\sigma$ is false under $\Gamma'$
	\item $(L; \pi)$ is undefined in $\Gamma$
	\item $L_1$ and $L_2$ become false by the decision, i.e.\
		$\neg L_1, \neg L_2 \in \mGnd(L; \pi)$
	\item $L_1 \ne L_2$
\end{itemize}
If no such $C$ exists, we say that \emph{the decision is not blocked in $\Gamma$}.
\end{defi}

\begin{example} Consider $\mDomain = \{a, b, c\}$, $\Gamma = \{(\neg Q(x,y); \top)^1\}$, and 
\[\text{N} = \{ C: \neg P(x) \lor \neg P(y) \lor Q(x,y), \dots \}\]
Then the decisions $(P(x); \top)$, $(P(x); x \ne c)$ are both blocked in $\Gamma$ by $C$, as 
witnessed by the ground instance $\neg P(a) \lor \neg P(b) \lor Q(a,b)$.
\end{example}

\noindent
We give our calculus as a set of rules over so-called \emph{states}, tuples of the form
\[(\Gamma; \text{N}; \text{U}; k; s)\]
where $\Gamma$ denotes the trail, $\text{N}$ the given clause set, $\text{U}$ the set of learned clauses,
$k$ a non-negative integer - unless terminating with \emph{Success} -, and $s$ a \emph{state indicator}. 
The latter can be $\top$, $\bot$, or a set of clause instances $\mGnd(C\sigma;\pi)$ given as $(C; \sigma; \pi)$.

$\top$ indicates the conflict search phase, if $k\geq 0$, or that $\Gamma$ defines a model for $\text{N}$, if $k = -1$. 
$\bot$ means the empty clause has been learned, i.e.\ 
the unsatisfiability of $\text{N}$ has been established.
Finally, an indicator of the form $(C; \sigma; \pi)$ represents a set of clause instances falsified by the current trail $\Gamma$, 
and indicates the conflict resolution phase of our calculus.

Our results extend to any derivation starting from a \emph{sound} state (see Definition~\ref{sndstatedef}). 
Here we propose the initial state 
\[(\epsilon; \text{N}; \emptyset; 0; \top)\]
where $\epsilon$ stands for the empty trail, and $\text{N}$ is the set of input clauses.

Next, we address a technical question regarding deduced literals and conflict resolution. 
It is often the case that for a clause $(C\lor L) \in \text{N}$ with $C\sigma$ implying $(L\sigma; \pi)$ for some $\sigma$, $\pi$ 
w.r.t.\
the current trail $\Gamma$, the involved substitution $\sigma$ substitutes variables not occurring in $L$. 
See Example~\ref{exProp} for a demonstration of this behavior.

Should we save only $(L\sigma; \pi)$ to the trail, we would lose this part of the assignment. 
However, during conflict resolution we need the exact clause instances responsible for the assignment. 
Therefore, to avoid recomputing the relevant substitutions, we save the \emph{constrained closure} $(L\cdot\sigma; \pi)$, where 
$L\cdot\sigma$ is the \emph{closure} representing $L\sigma$. 

This is an extension of the existing notation for the sake of clause learning. 
For all other purposes, $L\cdot\sigma$ is identified with $L\sigma$, and all definitions over constrained literals 
can be extended to constrained closures accordingly.
The literal $L$ is also considered to be a short-hand for $L\cdot\emptyset$.
We also note that in our calculus decisions are always considered having empty closures.

%
Finally, a short remark on the usage of the operations over constrained expressions:
Conjunction is used whenever we try to unify two constrained literals, 
e.g. during learning a new clause via resolution,
or finding candidates for propagation.
Difference is needed when we remove already defined literals ensuring 
that a new assignment only defines new values.
Emptiness is tested overall in the calculus to ensure that a new assignment 
indeed defines the value of at least one ground atom.

Below, we provide the rules of our calculus in a generic style as a state transition system, 
similarly to~\cite{NieuwenhuisOT04}. 
We note that in the rules $\pi_1, \pi_2$ is often used as a short-hand for $\pi_1 \land \pi_2$, if it is unambiguous.
Furthermore, blocking is considered only w.r.t.\ the current clause set $\text{N}\cup\text{U}$ in the rest of the paper. 
For further details on the applied strategy and technicalities, see Section 5 and Section 8.

\subsection{Rules for Conflict Search}
\input{src/s3-calc/ss3-csearch}

\subsection{Rules for Conflict Resolution}
\input{src/s3-calc/ss4-cresolution}
\subsection{Example}
\input{src/s3-calc/ss5-examples}

%% file: src/s3-calc/ss3-csearch.tex
%
\textbf{Propagate}
\[(\Gamma; \text{N}; \text{U}; k; \top) \Rightarrow
    (\Gamma, (L\cdot\sigma;\pi)^{C \lor L}; \text{N}; \text{U}; k; \top)\]
if $k \geq 0$, and for $(C \lor L) \in (\text{N} \cup \text{U})$, $\sigma$, and $\pi$
\begin{itemize}
	\item $(C\sigma; \pi)$ is false under $\Gamma$
	\item $(L\sigma; \pi)$ is undefined in $\Gamma$
	\item $(L\sigma; \pi)$ is not empty
\end{itemize}
This rule deduces new literals which have to be true under the current model assumption.
The conditions ensure that this step is sound and \emph{effective}, i.e.\ 
each ground literal defined by the 
added literal is indeed a consequence and at least one such literal exists.

\begin{example}\label{exProp}
Let $a,b \in \mDomain$, $\text{N}$ and $\text{U}$ arbitrary, $C_1, C_2 \in \text{N}$, and the current state
\[(\Gamma; \text{N}; \text{U}; 1; \top)\]
where
\[\Gamma = (P(x,x); \top)^{C_1}, (Q(a,x); \top)^{C_2}, (\neg P(x,y); (x,y) \ne (v,v))^1\]
Then, if $C = P(y,b) \lor \neg Q(x,y) \lor R(y)$ is a clause from $\text{N}$, \emph{Propagate} 
can be applied for $C$, and we might get the state
\[(\Gamma, (R(y)\cdot\{x \gets a\}; y \ne b)^C; \text{N}; \text{U}; 1; \top)\]
\end{example}
\textbf{Decide}
\[(\Gamma; \text{N}; \text{U}; k; \top) \Rightarrow
    (\Gamma, (L; \pi)^{k+1}; \text{N}; \text{U}; k+1; \top)\]
if $k \geq 0$, and for $L$, $\pi$
\begin{itemize}
	\item $(L; \pi)$ is undefined in $\Gamma$
	\item $(L; \pi)$ is not blocked in $\Gamma$
	\item $(L; \pi)$ is not empty
	\item $\exists (C \lor L') \in \text{N}$ such that $|L| \geq |L'|$, i.e.\ 
		$\exists \delta: L = L'\delta\text{, or }L = \neg L'\delta$
\end{itemize}
\emph{Decide} adds an assumption to $\Gamma$ which is not blocked by any of the clauses, and which 
is effective. 

We note that the last condition is optional, it does not influence any of our results.
This restriction allows earlier termination with \emph{Success} and keeps the calculus from defining irrelevant ground atoms. 
After terminating with \emph{Success}, every undefined ground atom can be considered having arbitrary truth-values, 
or simply \emph{false}, the way it is defined in $I_{\Gamma}$.

We also note that blocking only identifies one kind of immediate conflicts, we might still get to an 
outright conflict if it can be handled with factorization, see Example~\ref{exImmConf} below, and Lemma~\ref{lemmImmConfFact} for details.


\begin{example}\label{exImmConf}
Let $\mDomain = \{a, b, c\}$, $\Gamma = (P(x,x); \top)^{P(x,x)}, (Q(x,a); \top)^{Q(x,a)}$, and 
\[\text{N} = \{P(x,x), Q(x,a), \neg Q(x,y) \lor P(x,y) \lor P(x,y) \}\]
Then, the decision $(\neg P(x,y); (x,y) \ne (v,v))$ is not blocked, yet 
\[(\neg Q(x,y) \lor P(x,y) \lor P(x,y); \{y \gets a\}; x \ne a)\]
is false w.r.t.\
$\Gamma, (\neg P(x,y); (x,y) \ne (v,v))^1$.
We note that conflict resolution learns the clause $\neg Q(x,y) \lor P(x,y)$ from this conflict.
\end{example}
\noindent
We also note that whenever a decision is blocked, we can always pick a stricter unblocked decision, shown below.
\begin{prop}
For every blocked decision $(L; \pi)$ and blocking clause $C$, 
there is a decision $(L\sigma; \pi\sigma, \pi')$ for some $\sigma$, $\pi'$ such that 
it is not blocked by $C$ and it is not empty. 
\end{prop}
\noindent\startproof
It is easy to see that any ground literal from $\mGnd(L; \pi)$ satisfies this condition.
\finishproof\leaveabit
\textbf{Conflict}
\[(\Gamma; \text{N}; \text{U}; k; \top) \Rightarrow
    (\Gamma; \text{N}; \text{U}; k; (C; \sigma; \pi))\]
if $k \geq 0$, and for some $\bot \ne C \in (\text{N} \cup \text{U})$, $\sigma$, and $\pi$
\begin{itemize}
	\item $(C\sigma; \pi)$ is false under $\Gamma$
	\item $(C\sigma; \pi)$ is not empty
\end{itemize}
\emph{Conflict} identifies a set of clause instances contradicting the current model assumption. 
We also refer to this set as the \emph{conflict-set}.

\begin{example}
Let $\mDomain =\{a,b,c\}$, and 
\begin{align*}
	\text{N} =&~\{~\exClauseNo{1}: \neg P(c), \exClauseNo{2}: \neg P(x) \lor \neg P(y) \lor Q(x,y), \\
								   &~~~\exClauseNo{3}: \neg P(y) \lor \neg Q(a,y), \exClauseNo{4}: \neg Q(x,b)\lor\neg P(x)~\}\\
	\Gamma =&~(\neg P(c); \top)^{\exClauseNo{1}}, (P(x); x \ne c)^1, (\neg Q(a,y); y \ne c)^{\exClauseNo{3}}
\end{align*}
Then the following is a valid step:
\[(\Gamma; \text{N}; \emptyset; 1; \top) 
\stackrel{Conflict (\exClauseNo{2})}{\Rightarrow} 
(\Gamma; \text{N}; \emptyset; 1; (\neg P(x) \lor \neg P(y) \lor Q(x,y); \{x \gets a\}; y \ne c)) 
\]
\end{example}
\textbf{Success}
\[(\Gamma; \text{N}; \text{U}; k; \top) \Rightarrow
    (\Gamma; \text{N}; \text{U}; -1; \top)\]
if $k \geq 0$, and $I_{\Gamma} \models \text{N}$.\leaveabit\indent
We note that the last condition, $I_{\Gamma} \models \text{N}$, can be replaced by 
demanding that the rules \emph{Propagate}, \emph{Decide} and \emph{Conflict} are 
exhausted and $\bot \notin (\text{N}\cup\text{U})$.

From this it follows that each ground atom is defined and there is no falsified 
instance, i.e.\ 
every ground clause $C \in \mGnd(\text{N}\cup\text{U})$ is true w.r.t.\
the current trail.\leaveabit\noindent
\textbf{Failure}
\[(\Gamma; \text{N}; \text{U}; k; \top) \Rightarrow
    (\Gamma; \text{N}; \text{U}; 0; \bot)\]
if $\bot \in (\text{N} \cup \text{U})$.\leaveabit\indent
The two terminal rules correspond to the satisfiability and unsatisfiability of the clause set, respectively. 
Unsatisfiability is detected through learning the empty clause $\bot$.
%

%% file: src/s3-calc/ss4-cresolution.tex
%
\textbf{Skip}
\[(\Gamma, (L'\cdot\sigma'; \pi')^{C'}; \text{N}; \text{U}; k; (C; \sigma; \pi)) \Rightarrow
    (\Gamma; \text{N}; \text{U}; k; (C; \sigma; \pi))\]
if there is no $L \in C$ such that
\begin{itemize}
	\item $\exists \eta = \mMGU(L'\sigma', \neg L\sigma)$, and
	\item $(C\sigma\eta; \pi\eta, \pi'\eta)$ is not empty
\end{itemize}
\emph{Skip} drops the right-most literal from the trail during conflict resolution 
if it is not a decision and it does not contribute to the conflict, i.e.\ 
it does not touch any instance of the conflict-set.\leaveabit
%
\textbf{Resolve}
\begin{flushleft}$(\Gamma, (L'\cdot\sigma'; \pi')^{C' \lor L'}; \text{N}; \text{U}; k; (C \lor L; \sigma; \pi)) \Rightarrow$\end{flushleft}
  \begin{flushright} $(\Gamma, (L'\cdot\sigma'; \pi')^{C' \lor L'}; \text{N}; \text{U}; k; ((C \lor C')\eta_0; \sigma^*; \pi\eta, \pi'\eta))$\end{flushright}
if for $L'$, $\sigma$, $\pi'$ and $C' \lor L'$, and
\begin{itemize}
	\item $((C\lor L)\sigma; \pi)$ is not assertive, or $k = 0$
	\item $\exists \eta = \mMGU(L'\sigma', \neg L\sigma)$, and let
		\begin{itemize}
			\item $\eta_0 = \mMGU(L', \neg L)$
			\item $\sigma^*$ such that $\sigma\sigma'\eta = \eta_0\sigma^*$
		\end{itemize}
	\item $((C\lor L)\sigma\eta; \pi\eta, \pi'\eta)$ is not empty
\end{itemize}
We note that keeping $\sigma^*\rst{\mVar((C \lor C')\eta_0)}$ instead of $\sigma^*$ is enough
for the soundness of the rule and our calculus, as it contains all the relevant information.
Furthermore, the existence of $\eta$ implies the existence of $\eta_0$ and $\sigma^*$.

If the right-most literal in $\Gamma$ is not a decision and is involved in the conflict-set, we proceed with 
resolution. The conditions imply that there are corresponding ground inferences and the new conflict-set 
is not empty.

Note that dropping the used literal is not desired as the new conflict might still be resolvable with it.\leaveabit
\textbf{Factorize}
\[(\Gamma, \ell; \text{N}; \text{U}; k; (C \lor L_1 \lor L_2; \sigma; \pi)) \Rightarrow
    (\Gamma, \ell; \text{N}; \text{U}; k; ((C \lor L_1)\eta_0; \sigma^*; \pi\eta))\]
if $\ell = (L'\cdot\sigma'; \pi')^{\alpha}$ for some $L'$, $\sigma'$, $\pi'$, and annotation $\alpha$, and
\begin{itemize}
	\item $\exists \eta = \mMGU\{L_1\sigma, L_2\sigma, L'\sigma'\}$, and let
		\begin{itemize}
			\item $\eta_0 = \mMGU(L_1, L_2)$
			\item $\sigma^*$ such that $\sigma\eta = \eta_0\sigma^*$
		\end{itemize}
	\item $((C\lor L_1)\sigma\eta; \pi\eta, \pi'\eta)$ is not empty
\end{itemize}
Again, the existence of $\eta$ implies the existence of $\eta_0$ and the appropriate $\sigma^*$, and 
keeping $\sigma^*\restriction_{\mVar((C \lor C')\eta_0)}$ is sufficient.
We also note that $\alpha$ can be both a reason clause and a decision level.

\emph{Factorize} factorizes some of the conflicting ground clauses. 
As in the case of \emph{Resolve}, the used literal should not be dropped from the trail.\leaveabit
\textbf{Backjump}
\[(\Gamma_1, \Gamma_2; \text{N}; \text{U}; k; (C; \sigma; \pi)) \Rightarrow
    (\Gamma_1; \text{N}; \text{U} \cup \{C\}; k'; \top)\]
if $0 \leq k' \leq k$, $k' = \operatorname{lvl}(\Gamma_1)$, and one of the following condition-sets hold:
\begin{enumerate}[(1)]
	\item $k = 0$, and $C = \bot$, or
	\item $k > 0$, $(C\sigma; \pi)$ is assertive, and $C$ has no false instance under $\Gamma_1$, or
	\item $k > 0$, the right-most element of $\Gamma_2$ is the top-level decision, 
					$(C\sigma; \pi)$ is not assertive, \emph{Factorize} cannot be applied, and $C$ has no false instance under $\Gamma_1$
\end{enumerate}
It is clear that $k' = 0$ or $k' < k$ in case $(1)$ and $(2), (3)$, respectively. 

The optimal choice for $k'$ is the smallest level for which the learned clause can be used in \emph{Propagate}.
Such a $k'$ might not always exist for the learned clause $C$, largely due to the instances of $C$ not covered 
by $(C\sigma; \pi)$. In these cases the optimal choice for $k'$ is the largest level for which $C$ has no false instance.
For more details see Section 8.

In case $(1)$, we say that \emph{the empty clause $\bot$ is learned}. 
In case $(2)$, we say \emph{a new assertive clause} is learned, and in case $(3)$ \emph{a new blocking clause is learned}.

The latter clause is indeed blocking the last decision under some regularity conditions, see Lemma~\ref{lemmRegProps} for details.
We note that case $(3)$ can indeed occur as the following example demonstrates:

\begin{example}[Learning a blocking clause]\emph{~}\leaveabit
\noindent
Consider the clause set
\[\text{N} = \{ 
		\exClauseNo{1}:~R(x,x), 
		\exClauseNo{2}:~P(x) \lor \neg Q(x,y),
		\exClauseNo{3}:~R(x,y) \lor Q(x,y) \lor P(x) \lor P(y)
\}\]
and let $\Gamma = \Gamma', (\neg Q(x,y); \top)^{\exClauseNo{2}}$ with
\[\Gamma' = (R(x,x); \top)^{\exClauseNo{1}}, (\neg R(x,y); (x,y) \ne (v,v))^1, (\neg P(x); \top)^2\]
Then the following is a valid conflict resolution:
\[
(\Gamma; \text{N}; \emptyset; 2; (R(x,y) \lor Q(x,y) \lor P(x) \lor P(y); \emptyset; (x,y) \ne (v,v)))
\stackrel{Resolve}{\Rightarrow}
\]
\[
(\Gamma; \text{N}; \emptyset; 2; (R(x,y) \lor P(x) \lor P(x) \lor P(y); \emptyset; (x,y) \ne (v,v)))
\stackrel{Skip}{\Rightarrow}
\]
\[
(\Gamma'; \text{N}; \emptyset; 2; (R(x,y) \lor P(x) \lor P(x) \lor P(y); \emptyset; (x,y) \ne (v,v)))
\stackrel{Factorize}{\Rightarrow}
\]
\[
(\Gamma'; \text{N}; \emptyset; 2; (R(x,y) \lor P(x) \lor P(y); \emptyset; (x,y) \ne (v,v)))
\stackrel{Backjump(3)}{\Rightarrow}
\]
\[
((R(x,x); \top)^{\exClauseNo{1}}, (\neg R(x,y); (x,y) \ne (v,v))^1; \text{N}; \{R(x,y) \lor P(x) \lor P(y)\}; 1; \top)
\]
\end{example}

\begin{remark} We also wish to note that the current formulation of the calculus handles 
blocking decisions and learning blocking clauses asymmetrically in the following sense.\leaveabit
Let $\mDomain = \{a, b, c\}$, $\text{N} = \{P(x,x), Q(x,a), \neg Q(x,y) \lor P(x,y) \lor P(x',y) \}$, and 
\[\Gamma = (P(x,x); \top)^{P(x,x)}, (Q(x,a); \top)^{Q(x,a)}\]
Then the decision $(\neg P(x,y); (x,y) \ne (z,z))$ is blocked by $\neg Q(x,y) \lor P(x,y) \lor P(x',y)$. 
We could use factorization and learn $(\neg Q(x,y) \lor P(x,y))$, 
but instead we rather throw away the decision candidate and try another.\leaveabit
%
On the other hand, if in some regular run (see Definition~\ref{regRunDef}) a conflict state of the form
\[(\Gamma', \ell^{k}; \text{N'}; \text{U'}; k; (\neg Q(x,y) \lor P(x,y) \lor P(x',y); \{y \gets a\}; x \ne a \land x' \ne a))\]
with $\ell = (\neg P(x,y); (x,y) \ne (z,z))$ arises, we choose \emph{Factorize} over learning a blocking clause outright - there is indeed a blocking instance -, 
and learn the assertive and not-blocking $\neg Q(x,y) \lor P(x,y)$ in the end.
\end{remark}

%% file: src/s3-calc/ss5-examples.tex
%
\begin{example}

As an example, we present a derivation which constructs a model over $\mDomain = \{a,b,c\}$ for the clause set
\\[3 pt]\centerline{\begin{tabular}{rlll}
$\text{N} = \{$ & $\exClauseNo{1}: \neg P(c, x, x)$, $\exClauseNo{2}: \neg P(x, y, z) \lor \neg P(u, w, t) \lor Q(x, u),$ & \\
& $\exClauseNo{3}: \neg P(x, y, z) \lor \neg Q(a, x)$, $\exClauseNo{4}: \neg Q(x, b) \lor \neg P(x, y, z)$ & $\}$ \\
\end{tabular}}\\[3 pt]
The run below is by no means optimal - any sensible heuristic would choose the negative assignment for $P$ outright -, 
but it is a valid derivation, and serves well as a demonstration for the syntactic behavior.
\\[3 pt]\centerline{$
	(\epsilon; \text{N}; \emptyset; 0; \top)  
    \stackrel{Propagate}{\Rightarrow}
   ((\neg P(c, x, x); \top)^{\exClauseNo{1}}; \text{N}; \emptyset; 0; \top)  
	  \stackrel{Decide}{\Rightarrow}  
$}\\[3 pt]\centerline{$
	((\neg P(c, x, x); \top)^{\exClauseNo{1}}, (P(x, y, z); x \ne c)^1; \text{N}; \emptyset; 1; \top) 
	\stackrel{Propagate}{\Rightarrow}
$}\\[3 pt]\centerline{$
	((\neg P(c, x, x); \top)^{\exClauseNo{1}}, (P(x, y, z); x \ne c)^1, (\neg Q(a, x); x \ne c)^{\exClauseNo{3}}; \text{N}; \emptyset; 1; \top) 
	\stackrel{Conflict \exClauseNo{2}}{\Rightarrow} 
$}\\[3 pt]\centerline{$
(\dots; \text{N}; \emptyset; 1; (\neg P(x, y, z) \lor \neg P(u, w, t) \lor Q(x, u); \{x \gets a\}; u \ne c)) 
\stackrel{Resolve}{\Rightarrow} 
$}\\[3 pt]\centerline{$
(\dots; \text{N}; \emptyset; 1; (\neg P(a, y, z) \lor \neg P(u, w, t) \lor \neg P(u, y', z'); \emptyset ; u \ne c )) 
\stackrel{Skip}{\Rightarrow} 
$}\\[3 pt]\centerline{$
(\dots, (P(x, y, z); x \ne c)^1; \text{N}; \emptyset; 1; (\neg P(a, y, z) \lor \neg P(u, w, t) \lor \neg P(u, y', z'); \emptyset; u \ne c)) 
$}\\[3 pt]\centerline{$
\stackrel{Factorize}{\Rightarrow} 
(\dots, (P(x, y, z); x \ne c)^1; \text{N}; \emptyset; 1; (\neg P(a, y, z) \lor \neg P(u, w, t); \emptyset; u \ne c)) 
$}\\[3 pt]\centerline{$
\stackrel{Factorize}{\Rightarrow} 
((\neg P(c, x, x); \top)^{\exClauseNo{1}}, (P(x, y, z); x \ne c)^1; \text{N}; \emptyset; 1;(\neg P(a, y, z); \emptyset; \top)) 
$}\\[3 pt]
Let $\text{U}_1 = \{ \exClauseNo{5}:~\neg P(a, y, z)\}$.
\\[3 pt]\centerline{$
\stackrel{Backjump(2)}{\Rightarrow} 
((\neg P(c, x, x); \top)^{\exClauseNo{1}}; \text{N}; \text{U}_1; 0; \top) 
\stackrel{Propagate}{\Rightarrow}
$}\\[3 pt]\centerline{$
((\neg P(c, x, x); \top)^{\exClauseNo{1}}, (\neg P(a, y, z); \top)^{\exClauseNo{5}}; \text{N}; \text{U}_1; 0; \top) 
\stackrel{Decide}{\Rightarrow}
$}\\[3 pt]\centerline{$
((\neg P(c, x, x); \top)^{\exClauseNo{1}}, (\neg P(a, y, z); \top)^{\exClauseNo{5}}, (P(b, y, z); \top)^1; \text{N}; \text{U}_1; 1; \top) 
\stackrel{Propagate}{\Rightarrow}
$}\\[3 pt]\centerline{$
(\dots, (P(b, y, z); \top)^1, (Q(x, u)\cdot\sigma_1; \top)^{\exClauseNo{2}}; \text{N}; \text{U}_1; 1; \top) 
\stackrel{Conflict \exClauseNo{4}}{\Rightarrow}
$}\\[3 pt]
Where $\sigma_1 = \{x\gets b, u\gets b\}$.
\\[3 pt]\centerline{$
(\dots, (Q(x, u)\cdot\sigma_1; \top)^{\exClauseNo{2}}; \text{N}; \text{U}_1; 1; (\neg Q(x,b) \lor \neg P(x, y, z); \{x \gets b\}; \top)) 
\stackrel{Resolve}{\Rightarrow}
$}\\[3 pt]\centerline{$
(\dots; \text{N}; \text{U}_1; 1; (\neg P(x, y, z) \lor \neg P(x, y', z') \lor \neg P(b, w, t); \{x \gets b\}; \top)) 
\stackrel{Skip}{\Rightarrow}
$}\\[3 pt]\centerline{$
(\dots, (P(b, y, z); \top)^1; \text{N}; \text{U}_1; 1; (\neg P(x, y, z) \lor \neg P(x, y', z') \lor \neg P(b, w, t); \{x \gets b\}; \top)) 
$}\\[3 pt]\centerline{$
\stackrel{Factorize}{\Rightarrow}
(\dots, (P(b, y, z); \top)^1; \text{N}; \text{U}_1; 1; (\neg P(x, y, z) \lor \neg P(b, w, t); \{x \gets b\}; \top)) 
$}\\[3 pt]\centerline{$
\stackrel{Factorize}{\Rightarrow}
(\dots, (P(b, y, z); \top)^1; \text{N}; \text{U}_1; 1; (\neg P(b); \emptyset; \top)) 
\stackrel{Backjump(2)}{\Rightarrow}
$}\\[3 pt]\centerline{$
((\neg P(c, x ,x); \top)^{\exClauseNo{1}}, (\neg P(a, y, z); \top)^{\exClauseNo{5}}; \text{N}; \text{U}_1 \cup \{ \neg P(b, y, z) \}; 0; \top) 
\stackrel{Propagate}{\Rightarrow}
$}\\[3 pt]\centerline{$
((\neg P(c, x, x); \top)^{\exClauseNo{1}}, (\neg P(a, y, z); \top)^{\exClauseNo{5}}, (\neg P(b, y, z); \top)^{\exClauseNo{6}}; \text{N}; \text{U}_2; 0; \top) 
\stackrel{Decide}{\Rightarrow}
$}\\[3 pt]
Where $\text{U}_2 = \text{U}_1 \cup \{ \exClauseNo{6}:~\neg P(b, y, z) \}$.
\\[3 pt]\centerline{$
(\dots, (\neg P(b, y, z); \top)^{\exClauseNo{6}}, (\neg P(c, y, z); (y, z) \ne (v, v))^1; \text{N}; \text{U}_2; 1; \top) 
\stackrel{Decide}{\Rightarrow}
$}\\[3 pt]\centerline{$
(\dots, (\neg P(c, y, z); (y, z) \ne (v, v))^1, (Q(x, y); \top)^2; \text{N}; \text{U}_2; 2; \top) 
\stackrel{Success}{\Rightarrow}
$}\\[3 pt]\centerline{$
(\dots, (\neg P(c, y, z); (y, z) \ne (v, v))^1, (Q(x, y); \top)^2; \text{N}; \text{U}_2; -1; \top)
$}
\end{example}

%% file: src/s4-decproc/soundness.tex
%
\section{Soundness}

Now, we show soundness. The following state invariant defines a consistency notion for states.

\begin{defi}\label{sndstatedef}
A state $(\Gamma; \text{N}; \text{U}; k; s)$ is \emph{sound} if and only if the followings hold:
	\begin{enumerate}
		\item $\Gamma$ is a consistent sequence of constrained literals
		\item $\Gamma$ is \emph{well-formed}, i.e.\
			\begin{enumerate}
				\item if $k \geq 0$ then $\Gamma$ contains exactly $k$ decisions
				\item for each $i$ from $1, 2, \dots, k$, there is a unique $(L; \pi)^i\in\Gamma$
				\item the decisions occur in $\Gamma$ in the order of their levels
				\item for each decomposition $\Gamma = \Gamma_1, (L; \pi)^i, \Gamma_2$; $(L, \pi)^i$ satisfies the conditions of \emph{Decide} w.r.t.\
					$\Gamma_1$, $\text{N}$, and $\text{U}$
				\item for each decomposition $\Gamma = \Gamma_1, (L\cdot\sigma; \pi)^{C\lor L}, \Gamma_2$; 
								$(C\sigma; \pi)$ is false under $\Gamma_1$, 
								and $(L\sigma; \pi)$ satisfies the conditions for \emph{Propagate} w.r.t.\
								$\Gamma_1$ and $C\lor L$
			\end{enumerate}	
		\item $N \models U$
		\item $s = \bot$ implies $\bot \in \text{N}\cup\text{U}$
		\item $k = -1$ implies $I_{\Gamma}\models \text{N}$
		\item if $s = (C; \sigma; \pi)$ then 
			      $(C\sigma; \pi)$ is false under $\Gamma$, $N \models C$, and $(C\sigma; \pi)$ is not empty.
	\end{enumerate}
	A rule is called \emph{sound} iff it preserves the soundness of its left-hand side state.
\end{defi}

It is easy to see that the initial state $(\epsilon; \text{N}; \emptyset; 0; \top)$ is always sound.
Furthermore, soundness is an invariant, since each rule preserves this property, as proven below.

\begin{theo}\label{sndrulesTheo}
The rules of NRCL are sound.
\end{theo}
\noindent\startproof
The soundness of \emph{Propagate}, \emph{Decide}, \emph{Conflict}, and 
the terminal rules \emph{Failure} and \emph{Success} is straightforward 
to prove from the definitions themselves, and therefore, we entrust it to the reader.

In the case of \emph{Skip}, dropping the right-most literal $(L'\cdot\sigma'; \pi')^{C'}$ from $\Gamma$ does preserve 
the well-formedness and consistency properties of $\Gamma$. 
$\text{N} \models \text{U}$ remains unchanged and the rest 
of the conditions are irrelevant in this case, except for the last one.

Now, assume the last property does not hold after applying \emph{Skip}. It is only possible if some 
ground clause $C''$ from $\mGnd(C\sigma; \pi)$ were false under $\Gamma, (L'\cdot\sigma'; \pi')^{C'}$, but is undefined
under $\Gamma$. Thus, $(L'\sigma'; \pi')$ must have made it false, and therefore, for some $\delta$ and $L''\in C''$, 
$L'' = \neg L'\sigma'\delta$ and $\pi'\delta$ is true. 

Let $L$ be the literal in $C$ corresponding to $L''$. Then, the most general unifier $\eta$ of $\neg L\sigma$ and $L'\sigma'$ 
must exist and $C'' \in \mGnd(C\sigma\eta; \pi\eta, \pi'\eta)$, which is therefore not empty. This violates the 
preconditions of \emph{Skip}, a contradiction.

For \emph{Resolve}, it is enough to see that the new clause is a consequence of $\text{N}$, and the new 
state indicator $((C\lor C')\eta_0; \sigma^*; \pi\eta, \pi'\eta)$ is unsatisfiable under $\Gamma$, using the notations of 
the definition for \emph{Resolve}.

The first claim follows from the soundness of the left-hand side and from the soundness of resolution.
As for the second claim, we make the following observations:
\begin{itemize}
  \item $(C\lor C')\eta_0\sigma^* = (C' \lor C)\sigma\sigma'\eta$
	\item Each instance from $\mGnd(C'\sigma'\eta;\pi'\eta)$ is false under the current trail, as per the well-formedness conditions for derived literals.
	\item Each instance from $\mGnd(C\sigma\eta; \pi\eta)$ is false under the trail by the soundness of the left-hand side.
\end{itemize}
From these it follows that each ground clause from $\mGnd(C\lor C'; \sigma\sigma'\eta; \pi\eta, \pi'\eta)$ is false under the current trail.

The soundness of \emph{Factorize} can be proven analogously, and the proof for \emph{Backjump} is straightforward. 
We entrust them to the reader.
\finishproof\leaveabit
\noindent
Next, we define \emph{runs}, i.e.\ sound derivations in our calculus.

\begin{defi}
A \emph{run} (from a clause set $\text{N}$) is a sequence of states such that each subsequent state is derived with a rule from the previous one, and the initial state is a
sound state (with $\text{N}$ as the original clause set).
\end{defi}

A direct consequence of Theorem~\ref{sndrulesTheo} is that each state in a run is sound, and in particular, for each conflict resolution state 
$(\Gamma; \text{N}; \text{U}; k; (C; \sigma; \pi))$, each ground clause from $\mGnd(C\sigma; \pi)$ is false w.r.t.\ $\Gamma$.


\begin{theo}[Soundness]\label{sndRunTheo}
The calculus NRCL is sound, i.e.\ if a run terminates with the \emph{Failure}, or \emph{Success} rules, 
then the starting set $\text{N}$ is unsatisfiable, and satisfiable, respectively.
Furthermore, in the latter case the trail upon termination defines a model of $\text{N}$.
\end{theo}
\noindent\startproof
It follows immediately from the definitions and Theorem~\ref{sndrulesTheo}.
\finishproof
%

%% file: src/s4-decproc/regruns.tex
%
\section{Regular Runs}
In this section, we define a strategy for NRCL in the form of \emph{regular runs}, which is sufficient to prove both
non-redundant clause learning, and termination in the later sections. 

\begin{defi}\label{regStateDef}
A sound state $(\Gamma; \text{N}; \text{U}; k; s)$ is \emph{regular} iff the following hold:
	\begin{itemize}
		\item If $\Gamma = \Gamma', (L\cdot\sigma; \pi)^\alpha$, then no clause from $\text{N}\cup\text{U}$ is false w.r.t.\ $\Gamma'$.
		\item For all decomposition $\Gamma = \Gamma_1, (L; \pi)^i, \Gamma_2$ with decision $(L; \pi)^i$, \emph{Propagate} is exhausted w.r.t.\
			$\Gamma_1$ and $\text{N}\cup\text{U}$.
	\end{itemize}
\end{defi}
We note that the last assignment on the trail might still make some clauses false, and the initial state $(\epsilon; \text{N}; \emptyset; 0; \top)$ is always regular.
\begin{defi}\label{regRunDef}
We call a run \emph{regular} iff the following holds:
	\begin{itemize}
		\item The starting state is regular.
		\item During conflict search, rules are always applied in this order exhaustively: terminal rules, \emph{Conflict}, \emph{Propagate}, \emph{Decide}.
		(Or \emph{Failure}, \emph{Conflict}, \emph{Propagate}, \emph{Decide}, \emph{Success}, if we test success through exhausted conflict search.)
		\item In conflict resolution \emph{Backjump} is always applied as soon as possible, and it backtracks to a regular state.
	\end{itemize}
\end{defi}
\begin{lemm} Regular runs preserve regularity, i.e.\ every state in a regular run is regular.
\end{lemm}
\startproof
It follows from the definitions, we only note that backjumping to a state which is regular w.r.t.\
the new learned clause set as well is always possible.
If nothing else, the empty trail is always a valid choice. 
\finishproof\leaveabit
\noindent
The backtrack-level proposed in the proof above is not practical, of course.
For more details on a more accurate backjumping to a regular state see Section 8.\leaveabit\noindent
Below, we show some useful properties of regular runs.

\begin{lemm}\label{lemmRegProps} In a regular run the following hold:
  \begin{enumerate}[(1)]
		\item For any deduced literal $(L\cdot\sigma; \pi)^{C \lor L}$ of level $k$ on the trail with $k > 0$, each ground clause in $\mGnd((C \lor L)\sigma; \pi)$ contains at least two literals of level $k$.
		\item If $(C; \sigma; \pi)$ represents false clauses in some conflict state, then each ground clause in $\mGnd(C\sigma; \pi)$ contains at least two top-level literals, if 
		 the state is the result of an application of \emph{Conflict}, and at least one top-level literal otherwise.
		\item If a clause $C$ is learned according to the case \emph{Backjump}-$(3)$, then it blocks the former top-level decision.
  \end{enumerate}
\end{lemm}
\noindent
\startproof
First, assume $(L\cdot\sigma; \pi)$ is a deduced literal and it was implied by $(C\lor L; \sigma; \pi)$ w.r.t.\ 
$\Gamma$ which was the current trail before the corresponding application of \emph{Propagate}. 

Let $k$ be the level of the right-most decision in $\Gamma$, and $C' \lor L'$ a ground clause from $\mGnd((C\lor L)\sigma; \pi)$ 
such that $L'$ corresponds to $L$. 
Then $L'$ is of level $k$, of course. 

Furthermore, if no other literal in $C'$ is of level $k$, $C' \lor L'$ would have implied $L'$ 
before the last decision, which contradicts the exhaustive application of \emph{Propagate}.
Thus, $C' \lor L'$ must contain at least two literals of level $k$.

Second, since conflicts are found immediately, any conflicting non-empty ground clause $C'$ must contain at least one top-level literal. 
A conflicting ground clause with a single top-level literal, however, would contradict the exhaustive application of \emph{Propagate}.
Thus, after applying \emph{Conflict}, all ground clause in the conflict-set contains at least two top-level literals. 
It only remains to show that the rules \emph{Resolve}, \emph{Skip}, and \emph{Factorize} 
preserve the weaker property of having at least one top-level literals. Obviously, e.g.\ \emph{Factorize} can break the stronger property.

We only prove this for \emph{Resolve}, the rest can be shown similarly. 
Assume that at an application of \emph{Resolve} $(L'\cdot\sigma'; \pi')^{C'\lor L'}$  is the involved deduced literal, 
$(C \lor \neg L; \sigma; \pi)$ represents the false clauses before, and
$((C \lor C')\eta_0; \sigma^*; \pi\eta, \pi'\eta)$ after applying the rule, where 
$\eta = \mMGU(L'\sigma', L\sigma)$,
$\eta_0 = \mMGU(L', L)$, and $\sigma^*$ such that $\eta_0\sigma^* = \sigma\sigma'\eta$.

It is easy to see that for every ground clause 
\[(C_0 \lor C'_0) \in \mGnd((C \lor C')\sigma\sigma'\eta; \pi\eta, \pi'\eta)\]
there are corresponding ground clauses $(C'_0 \lor L'_0) \in \mGnd((C'\lor L')\sigma'\eta; \pi'\eta)$ and 
$(C_0 \lor \neg L_0) \in \mGnd((C \lor \neg L)\sigma\eta; \pi\eta)$ 
whose resolvent is exactly $(C_0 \lor C'_0)$, 
and $L_0$, $L'_0$ correspond to $L$ and $L'$, respectively, and $L_0 = L'_0$.

Then, by the first claim of this lemma, $C'_0$ must contain at least one top-level literals, and so does $C_0 \lor C'_0$.

Finally, assume $C$ is learned when case $(3)$ of \emph{Backjump} is applied to the state
\[(\Gamma, (L; \pi)^k; \text{N}; \text{U}; k; (C; \sigma; \pi))\]
Now, let $(C' \lor L'_1 \lor \dots \lor L'_s) \in \mGnd(C\sigma; \pi)$ an arbitrary ground clause, 
where $L'_1$, \dots, $L'_s$ denotes the top-level literals of the clause.

By (2), $s \geq 1$, and, since $(C\sigma; \pi)$ has no assertive clause, even $s \geq 2$ must hold. 
We also know that \emph{Factorize} was not applicable, thus, for any $i \ne j$ from $1, \dots, s$, 
$L_i \ne L_j$ holds.
Thus, $C$ blocks the decision $(L; \pi)$ w.r.t.\
$\Gamma$, as witnessed by the ground clause above.
\finishproof\leaveabit
\indent
It can be also shown that if there is an immediate conflict after a decision in a regular run, \emph{Factorize} is applied next.

\begin{lemm}\label{lemmImmConfFact} Assume 
\[
\stackrel{Decide}{\Rightarrow}
(\Gamma, (L; \pi)^{k}; \text{N}; \text{U}; k; \top)
\stackrel{Conflict}{\Rightarrow}
(\Gamma, (L; \pi)^{k}; \text{N}; \text{U}; k; (C; \sigma'; \pi'))
\]
is a valid subderivation in a regular run. 
Then \emph{Factorize}, and only \emph{Factorize}, is applicable to the conflict state $(\Gamma, (L; \pi)^{k}; \text{N}; \text{U}; k; (C; \sigma'; \pi'))$.
\end{lemm}
\noindent\startproof
Obviously, \emph{Resolve} and \emph{Skip} cannot be applied. 
Furthermore, if case \emph{Backjump}-$(3)$ were applicable, 
there would be a ground clause in $\mGnd(C\sigma';\pi')$ 
blocking the last decision, a contradiction. 

Also, there cannot be any ground clause in $\mGnd(C\sigma';\pi')$ with a single top-level literal, since otherwise \emph{Propagate} would not have been applied exhaustively 
before the decision. And $C = \bot$ cannot hold either, as otherwise \emph{Failure} should have been applied earlier. 
Thus, the other cases of \emph{Backjump} do not apply either.

Finally, let $C_0$ a ground clause from $\mGnd(C\sigma';\pi')$. 
This clause exists, and must contain at least two top-level literals, see Lemma~\ref{lemmRegProps}(2). 
These literals are falsified by the last decision, and do not block the decision. 

Let $L_0, K_0$ two such literals and $C_0 = C_0' \lor L_0 \lor K_0$. 
Then these literals are equal, and the corresponding literals $L_1$, $K_1$ in $C\sigma$ 
are unifiable. 

Then \emph{Factorize} is applicable unifying $L_1$ and $K_1$, and $C_0' \lor L_0$ can be used to prove the non-emptiness condition.
\finishproof
%

%% file: src/s4-decproc/redundancy.tex
%
\section{Redundancy}

We define redundancy w.r.t.\ 
the induced ordering $<_{\Gamma}$ in the standard way:
\begin{defi}\label{a-redundancyDefs}
A ground clause $C$ is \emph{redundant w.r.t.\
a ground clause set $\text{N}$ (and $<_{\Gamma}$)} iff
\[C \in \text{N}, \text{or }\exists S \subseteq \text{N}^{~<_{\Gamma} C}: S \models C\]
A first-order clause $C$ is \emph{redundant w.r.t.\
the first-order clause set $\text{N}$ (and $<_{\Gamma}$)} iff
\[\forall C' \in \mGnd(C): C'\text{ is redundant w.r.t.~}\mGnd(\text{N})\]
If redundancy does not hold, we call the corresponding clause \emph{non-redundant}, or \emph{irredundant}.
\end{defi}
%
%
\subsection{Learning Non-Redundant Clauses}
First, we show that each learned clause is non-redundant w.r.t.\
the current clause set and induced ordering.

The most important consequence of this theorem that checking the learned clauses for 
redundancy criterions which are independent of the concrete induced orderings can be spared. 

Such admissible criterions include subsumption, subsumption resolution and tautologies, as it is shown in the next subsection. 

\begin{theo}[Non-redundant Clause Learning]\label{freshnessTheo}
Let $\Gamma$ denote the trail at a conflict in a regular run, $<_{\Gamma}$ the induced ordering, and 
assume the clause $C$ is learned via the \emph{Backjump} rule, and 
let $\text{N}$ and $\text{U}$ be the starting clause set and the set of learned clauses before the conflict, respectively.
\leaveabit
Then, $C$ is not redundant w.r.t.\
$\text{N}\cup\text{U}$ and $<_{\Gamma}$.
\end{theo}
\noindent
\startproof
Assume the first and last state in conflict resolution is 
\[(\Gamma; \text{N}; \text{U}; k; (C_0;\sigma_0;\pi_0)) \Rightarrow^* (\Gamma'; \text{N}; \text{U}; k; (C; \sigma_1; \pi_1))\]
By soundness, $\text{N}\cup\text{U} \models C$ 
and each $C' \in \mGnd(C\sigma_1; \pi_1)$ is false w.r.t.\
both $\Gamma'$ and $\Gamma$.

Now let $C' \in \mGnd(C\sigma_1; \pi_1)$ and assume there is an 
$S \subset \mGnd(\text{N}\cup\text{U})$ such that $S \models C'$ and $S <_{\Gamma} C'$.
Because of $S <_{\Gamma} C'$, each $C'' \in S$ has a defined truth-value w.r.t.\
$\Gamma$. If all $C'' \in S$ is true, then, by $S \models C'$, so is $C'$, a contradiction.

Thus, let $C'' \in S$ arbitrary such that $C''$ is false under $\Gamma$.
We distinguish two cases whether $\Gamma'$ is a strict subset of $\Gamma$, or equal to it.

First, if $\Gamma' \ne \Gamma$, at least one \emph{Skip} had to be used,  and $C'$ contains no literal covered by the 
right-most literal of $\Gamma$. Neither does $C''$, since $C'' <_{\Gamma} C'$. But then, $C''$ has a defined truth-value and it 
can only be true, as otherwise an earlier conflict detection would have been possible. A contradiction.

Second, assume $\Gamma' = \Gamma$. 
If the right-most literal is a decision, no false clause from $\mGnd(C_0\sigma_0;\pi_0)$ blocks this decision, and  
\emph{Factorize} had to be applied several times followed by an application of case $(2)$ of \emph{Backjump}. 
(See also Lemma~\ref{lemmImmConfFact} on immediate conflicts.)

Let now $C'$ such that it contains only a single top-level literal. 
Since case $(2)$ of \emph{Backjump} was used, such a clause from $\mGnd(C\sigma_1; \pi_1)$ 
exists. Since $C''$ is false and it was undefined before, 
it contains some top-level literals. 

Since it was not a subject of \emph{Propagate} before the right-most decision, 
it has to contain at least two such literals. 
But $C'$ contains only one, and therefore $\mDef(C') <_p \mDef(C'')$ and $C' <_{\Gamma} C''$ must hold, a contradiction.

Finally, if $\Gamma = \Gamma'$ and the right-most literal is not a decision, the last rule had to be \emph{Backjump} (case $1$ or $2$),
and the same argumentation holds: 
If an assertive clause is learned, let $C'$ an instance from $\mGnd(C\sigma_1; \pi_1)$ such that it contains only a single top-level literal. However, $C''$ 
must contain at least two top-level literals, which again leads to $C' <_{\Gamma} C''$, a contradiction.
If $C = \bot$ is learned, it is smaller than any non-empty clause, and due to regularity, $\bot$ is a newly learned clause.
\finishproof
%
%
\subsection{Admissible Redundancies}
Next, we show that the classic redundancy criterions \emph{tautology}, \emph{strict subsumption}, and \emph{subsumption resolution} 
are \emph{admissible redundancies in NRCL}, i.e.\ 
the clauses these rules remove are indeed redundant w.r.t.\
any induced ordering.

\begin{prop}[Tautology] Let $C$ a clause and $\text{N}$ an arbitrary clause set.
\[\text{If $\models C$ holds, then $C$ is redundant w.r.t.\
$\text{N}$.}\]
\end{prop}
\noindent\startproof
Clearly, any ground instance of $C$ is a ground tautology and redundant, since it follows from the empty set which "contains" only smaller clauses.
\finishproof\leaveabit
\indent
Furthermore,  we also note that removing $C$ has no effect on any run of the calculus, 
since no instance of $C$ can be ever a conflict clause or imply an assignment.

\begin{prop}[Strict Subsumption]\label{propSubs}
Let $C$, $D$ be clauses, $\sigma$ a substitution, and $\text{N}$ a set of clauses. 
\[\text{If $C\sigma \subset D$, then $D$ is redundant w.r.t.\
$\text{N} \cup \{C\}$.}\]
\end{prop}
\noindent\startproof
Let $D\delta$ be a ground instance of $D$. Then $C\sigma\delta \subset D\delta$ and $C\sigma\delta <_{\Gamma} D\delta$ holds, for any induced ordering $<_{\Gamma}$. 
The latter holds, because $\mDef(C\sigma\delta) <_p \mDef(D\delta)$ holds in the abstract ordering.

Thus, $D\delta$ is redundant w.r.t.\
$\{C\sigma\delta\}$, and so is $D$ w.r.t.\
$\text{N} \cup \{C\}$, and strict subsumption is admissible.
\finishproof\leaveabit
\indent
Similarly to tautology, removing a subsumed clause has little effect on the calculus, since whenever the subsumed clause is  a conflict or a reason clause, the subsuming clause is either a
conflict clause or implying the same assignment as well.

\begin{prop}[Subsumption Resolution] 
Let $C$, $D$ clauses, $L$ a literal, $\sigma$ a substitution, and $\text{N}$ a clause set.
\[\text{If $C\sigma \subseteq D$ holds, then $D \lor \neg L\sigma$ is redundant w.r.t.\
$\text{N} \cup \{C \lor L, D\}$.}\]
\end{prop}
\noindent\startproof
Redundancy clearly holds as $D$ subsumes $D \lor \neg L\sigma$. 
Furthermore, we note that exchanging $D \lor \neg L\sigma$ with $D$ in the presence of $C \lor L$ is a sound step.
Thus, subsumption resolution as a rule for reducing a clause is admissible.
\finishproof
%
%

%% file: src/s4-decproc/decision-procedure.tex
%
\section{Termination and Completeness}

Just as most related calculi, NRCL is a decision procedure for {\mEPRs} as well, under the regularity conditions of Definition~\ref{regRunDef}.
Below, we show that regular runs never get stuck and eventually terminate.

%
%
\begin{prop}\label{noStuckProp}
A regular run is never stuck,  i.e.\ 
it terminates with the terminal rules, or one of the other rules is applicable.
\end{prop}
\noindent\startproof
It is enough to show that, unless we already terminated, a rule is always applicable. First, we show that conflict search cannot get stuck.

If $\bot$ is already in one of the clause sets, \emph{Failure} is applicable and we terminate. Thus, w.l.o.g. assume $\bot \notin \text{N}\cup\text{U}$.

Assume $\Gamma$ is total, i.e.\ 
defines each ground atom. Then $I_{\Gamma}$ defines all ground atom occurring in $\mGnd(\text{N})$,
 and it either satisfies $\text{N}$ or there is a 
false ground clause from $\mGnd(\text{N}\cup\text{U})$. 
In the first case, \emph{Success} is applicable, and \emph{Conflict} in the second case.

If $\Gamma$ is not total, and some undefined ground literal is implied by some ground clause, \emph{Propagate} is applicable. Otherwise, 
if no ground literal is implied and there is an undefined ground atom, we can always apply \emph{Decide}. 
We note that decisions which define only a single ground atom are never blocked.

Second, assume we are resolving a conflict, i.e.\ 
the state indicator is $(C; \sigma; \pi)$ for some $C$, $\sigma$, and $\pi$. 
If the top literal in $\Gamma$ is a decision and if $(C\sigma; \pi)$ is assertive, then \emph{Backjump} is applicable.
If it is not assertive, then either \emph{Factorize}, or case $(3)$ of \emph{Backjump} is applicable.

If the top literal is a deduced literal, 
and neither does $C = \bot$ hold, nor is $(C\sigma; \pi)$ assertive - in these cases \emph{Backjump} is applicable -, 
then we check the conditions of \emph{Skip}.
If \emph{Skip} is not applicable, it satisfies the conditions of \emph{Resolve}. 
Therefore, either \emph{Skip}, \emph{Factorize}, \emph{Resolve} must be applicable in this case.
\finishproof\leaveabit
\noindent
We show termination through a series of lemmas. First, we prove that both conflict search and conflict resolution always terminate:

%
%
\begin{lemm}\label{csTerminatesTheo} Assume $\text{N}$, $\Sigma$ and $\mDomain$ are all finite.
Then, a conflict search phase of a regular run always terminates, i.e.\ 
leads either to a conflict or to termination.
\end{lemm}
\noindent\startproof
By the finiteness of $\Sigma$, we know that $\mathcal{A}_{\Sigma}$ is also finite. 
Since a regular run is a series of sound steps, we also know that each application of \emph{Propagate} and \emph{Decide} 
defines at least one formerly undefined ground atom. 

Thus, a regular run eventually exhausts these rules, and, since it cannot get stuck by Proposition~\ref{noStuckProp}, one of the rules 
\emph{Failure}, \emph{Success}, or \emph{Conflict} has to be applied. And thereby, the conflict search phase in question ends.
\finishproof

%
%
\begin{lemm}\label{crTerminatesTheo} Assume $\text{N}$, $\Sigma$ and $\mDomain$ are all finite.
Then, a conflict resolution phase of a regular run always terminates, i.e.\ 
leads to the application of 
\emph{Backjump} in finitely many steps.
\end{lemm}
\noindent\startproof
Let us assign to each intermediate state $(\Gamma; \text{N}; \text{U}; k; (C; \sigma; \pi))$ in a conflict resolution 
the tuple $(\#(\Gamma); \mGnd(C\sigma; \pi))$ as a measure, where $\#(\Gamma)$ denotes 
the number of elements in $\Gamma$. 

Let us order these tuples with the lexicographical ordering $<_{lex}$ based on the canonical ordering over non-negative integers and $<_0$ 
where $<_0$ denotes both the ordering induced by the trail 
after finding the conflict, and its multiset extension. 
This ordering is well-founded.

We note that conflict resolution cannot get stuck, see Proposition~\ref{noStuckProp}. 
Therefore, it is enough to show that each application of the rules \emph{Skip}, \emph{Resolve}, and \emph{Factorize} strictly 
decreases our measure.

\emph{Skip} strictly decreases the size of $\Gamma$, and therefore our measure as well. 
In the case of \emph{Resolve} and \emph{Factorize}, 
it is enough to give a function satisfying the conditions of Proposition~\ref{orderPropProp} 
between the false instances on the two sides, i.e.\ 
a function $\gamma$
which assigns ground clauses from the right-hand side conflict-set to 
larger ground clauses from the left-hand conflict-set.

First, assume we apply \emph{Resolve} to the state 
	\[(\Gamma, (L'\cdot\sigma'; \pi')^{C' \lor L'}; \text{N}; \text{U}; k; (C \lor L; \sigma; \pi))\]
and we get
  \[(\Gamma, (L'\cdot\sigma'; \pi')^{C' \lor L'}; \text{N}; \text{U}; k; ((C \lor C')\eta_0; \sigma^*; \pi\eta, \pi'\eta))\]
where $\eta = \mMGU(L'\sigma', \neg L\sigma)$, 
$\eta_0 = \mMGU(L', \neg L)$, and 
$\sigma^*$ such that $\eta_0\sigma^* = \sigma\sigma'\eta$. 
For the sake of readability, let us introduce the symbols $\alpha = \sigma\sigma'\eta$ and $\pi^* = \pi\eta, \pi'\eta$.

Now, let $\beta$ be a grounding substitution such that $(C\lor C')\alpha\beta \in \mGnd( (C \lor C')\alpha; \pi^*)$. Since it was derived via resolution, 
there is a corresponding valid ground resolution step with premises 
	\begin{itemize}
		\item $C_1  \lor L_1 \in \mGnd((C \lor L)\alpha; \pi\eta)$
		\item $C_2 \lor L_2 \in \mGnd((C' \lor L')\alpha; \pi'\eta)$
	\end{itemize}
where we assume $L_1$ and $L_2$ are the literals corresponding to $L$ and $L'$, respectively. 
Since we apply resolution, we also know that $L_1 = \neg L_2$, and $(C\lor C')\alpha\beta = C_1 \lor C_2$.

By the definition of sound states and \emph{Propagate}, 
we know that $C_2$ contains only literals which were defined before the last assignment, 
and thus, $C_2 <_0 (\neg) L_2$, and therefore $C_2 <_0 L_1$. 
Then, $C_1 \lor C_2 <_0 C_1 \lor L_1$ must hold, 
and thus, we shall define $\gamma(C_1 \lor C_2)$ as $C_1 \lor L_1$.

Since $\gamma$ can be defined over the whole $\mGnd((C \lor C')\alpha; \pi^*)$ and $\mGnd((C \lor L)\alpha; \pi\eta)$ is a subset of 
$\mGnd((C \lor L)\sigma; \pi)$, we can apply Proposition~\ref{orderPropProp}, and we get 
\[\mGnd((C \lor L)\sigma; \pi) >_{0} \mGnd((C \lor C')\alpha; \pi^*)\]
and our measure strictly decreases, as the size of the trail is unchanged.
The proof for \emph{Factorize} is analogous.
\finishproof\leaveabit
\noindent
Next, we show that only finitely many new clauses can be learned thanks to our non-redundancy results in Theorem~\ref{freshnessTheo}.
%
%
\begin{lemm}\label{finiteLearningTheo}
If $\text{N}$, $\Sigma$ and $\mDomain$ are finite, a regular run can only learn finitely many new clauses.
\end{lemm}
\noindent\startproof
We use Higman's Lemma~\cite{Higman52} to prove this claim.
The lemma states that given an infinite sequence $w_1, w_2, \dots$ of words over a finite alphabet, there is always an index $i$ and a subsequent index $j$ such 
that the word $w_i$ is \emph{embedded} into $w_j$, i.e.\ after deleting some letters from $w_j$ we can get $w_i$.

Now, consider $\mathcal{A}_{\Sigma}$. Since $\Sigma$ and $\mDomain$ are finite, both the set of ground atoms and ground literals over $\Sigma$ and $\mDomain$ are finite.
The latter serves as the finite alphabet for our proof.

Since every learned clause is non-redundant at the time they are learned, by Theorem~\ref{freshnessTheo}, 
we can assign a non-redundant ground instance to any learned clause, by the definition of redundancy.

Assume we learn infinitely many clauses, and let us consider the assigned ground clauses $C_1, C_2, \dots$, where
$C_1$ is assigned to the clause learned at the first conflict, $C_2$ to the clause learned at the second, and so on.

Now, take any term ordering $>$, order the literals of the clauses, and assign this ordered sequence of literals to each clause. 
Let us denote this word over the alphabet of ground literals by $w(C)$ for every ground clause $C$.

Then, by Higman's Lemma, there are indices $i < j$ such that $w(C_i)$ is embedded in $w(C_j)$.
But it means that $C_i \subseteq C_j$, i.e.\ $C_j$ is strictly subsumed by or equal to $C_i$.

The admissibility of strict subsumption was proven in Proposition~\ref{propSubs}, and clearly 
an already present ground clause cannot be non-redundant either, for any induced ordering.
Thus, $C_j$ cannot be redundant at the $j$th conflict, a contradiction.
\finishproof\leaveabit
\noindent
Finally, we show termination, and state the main result as a corollary.
%
%
\begin{theo}[Termination]\label{termRunTheo}
A regular run always terminates if $\text{N}$, $\Sigma$ and $\mDomain$ are finite.
\end{theo}
\noindent\startproof 
First, we note that a run can be seen as a series of conflict search and conflict resolution phases, which ideally ends with a terminal rule.
By Lemma~\ref{csTerminatesTheo}, Lemma~\ref{crTerminatesTheo}, and Proposition~\ref{noStuckProp}, 
we know that each phase ends after finitely many steps without getting stuck. 

Thus, an infinite run must be an infinite series of conflict search and resolution sequences. Since each conflict resolution ends with 
\emph{Backjump}, it would imply that infinitely many new clauses are learned. But it contradicts Lemma~\ref{finiteLearningTheo}. 
\finishproof

%
%
\begin{coro}[Decision Procedure]
Regular runs provide a decision procedure for the {\mEPR} fragment if $\text{N}$, $\Sigma$ and $\mDomain$ are finite.

I.e.\ every regular run terminates after finitely many steps with \emph{Failure}, or \emph{Success}, for an unsatisfiable, or satisfiable 
clause set $\text{N}$, respectively.
\end{coro}
\noindent\startproof It follows from Proposition~\ref{noStuckProp} and the Theorems~\ref{sndRunTheo} and~\ref{termRunTheo}.
\finishproof
%

%% file: src/s8-tech/technicalities.tex
%
\section{Towards Implementation}
This far we considered mostly our calculus in an abstract fashion, and it is enough to establish 
the results of the previous chapters.

Here, we elaborate some details regarding the constraints, 
and refine some steps to bring NRCL closer to practical application. 
In particular, we provide an abstract algorithm for exhaustive propagation, 
to highlight some important difficulties and expensive steps in the calculus.

However, this section does not aim to provide a complete abstract algorithm for regular runs, we 
only briefly address some challenges and propose some solutions and approaches, 
which provides us a starting point for later implementation and experimentation.
%
\subsection{Free Variables}
\input{src/s8-tech/p0-freevars}
\subsection{Indexing Scheme}
\input{src/s8-tech/p1-wls}
\subsection{Finding Candidates}
\input{src/s8-tech/p2-cand}
\subsection{Exhaustive Propagation}
\input{src/s8-tech/p3-prop}
%
\subsection{Picking the Next Decision}
\input{src/s8-tech/ss2a-decisions}
%
\subsection{Ranking Literals}
\input{src/s8-tech/ss2-vsids}
%
\subsection{Clause Learning and Backjumping}
\input{src/s8-tech/ss3-cl}
%

%% file: src/s8-tech/p0-freevars.tex
%
\newcommand\restr[2]{\ensuremath{\left.#1\right|_{#2}}}
The definition of normal form for constrained literals demands the left-hand side of a constraint to contain only variables 
occurring in the constrained literal. 
Our calculus derives new assignments, i.e.\ new constrained literals for $\Gamma$, by applying 
resolution between the literals in $\Gamma$ and the clauses in $\text{N} \cup \text{U}$. 

However, even after normalization, the resulting candidate $(L\cdot\sigma; \pi)^C$ might contain
free left-hand side variables, i.e.\ variables which occur in the reason clause instance $C\sigma$, 
and still occur in $\mLVar(\pi)$, 
but do not occur in $L\sigma$. 
The following example demonstrates this behavior.

\begin{example}\label{freeVarEx}
Let us take
\[ \text{N} = \{
\exClauseNo{1}:~\neg Q(x,x), 
\exClauseNo{2}:~\neg Q(x,y) \lor \neg Q(x,z) \lor P(y,z)
\}
\]
And assume that after an application of \emph{Propagate} and \emph{Decide} we get the trail
\[
\Gamma = (\neg Q(x,x); \top)^{\exClauseNo{1}}, (Q(x,y); (x,y) \ne (v,v))^1
\]
Now, applying \emph{Propagate} between $\Gamma$ and the clause $\exClauseNo{2}$, we get the constrained literal
\[(P(y,z); (x,y) \ne (v,v) \land (x,z) \ne (w,w))\]
Over $\mDomain = \{a,b\}$, this constraint is satisfiable, 
the cover-set is $\{P(a,a), P(b,b)\}$, and after eliminating the free variable $x$ we get the constrained literals
\[(P(y,z); y \ne a \land z \ne a)\text{ and }(P(y,z); y \ne b \land z \ne b)\]
\end{example}

Semantically, these variables are to be treated as existential variables, of course. 
These variables cause two problems. 

First, in the presence of these existentially handled variables 
our constrained literal set for difference defined in Lemma~\ref{diffLemma2} is no longer valid. 
In particular disjointness is no longer guaranteed. 

A simple way to overcome this issue is to split the resulting literal into a set of literals by instantiating 
the free left-hand side variables with relevant constants, as seen in the example above. 
This elimination procedure results in a set of not necessarily disjoint constrained literals.

Second, while eliminating these variables is a solution, we still need to store the instantiating assignments.
This information is used 
when applying the rules \emph{Resolve} and \emph{Factorize} during conflict resolution.
This is already accomplished through using \emph{closures} as introduced in Section 3.

%% file: src/s8-tech/p1-wls.tex
%
In the propositional setting, the watched literal scheme watches two literals in every non-unit clauses. 
These literals are assumed to be true or undefined under the current model assumption, or all literals but a single watched literal are false in the clause.

Whenever a new assignment makes a watched literal false, 
we attempt to find a new non-false literal. 
If it is not possible, the other watched literal is propagated resulting either in a new assignment or a new conflict clause.

This scheme enables efficient propagation at small computational costs as it cuts back the number of clauses we have to consider
after a new assignment and requires no additional bookkeeping during backtrack.

When lifting the scheme, we have to keep in mind that manipulating our constraints is more expensive.
Therefore, a direct lifting of the technique by exactly maintaining which literals are watched in the different instances of a clause would be 
too expensive for our purposes.

Here, we propose a lightweight approach which uses two levels of indexing the literals of the current clause set.
Every clause is indexed by one of these levels, but not both.

The first level attempts to mimic the two-watched-literal scheme, and indexes only two literals in the clauses. 
We can choose the interpretation of watching a literal $L$ as an approximation of \emph{cannot be false} by selecting one of the following:
\begin{itemize}
	\item $\nexists (L'\cdot\sigma; \pi)^\alpha \in \Gamma: \exists\mMGU(\neg L'\sigma, L)$
	\item $\nexists (L'\cdot\sigma; \pi)^\alpha \in \Gamma: \exists\delta = \mMGU(\neg L'\sigma, L)\text{ and } \pi\delta \ne \bot$
	\item $\nexists (L'\cdot\sigma; \pi)^\alpha \in \Gamma: \exists\delta = \mMGU(\neg L'\sigma, L)\text{ and } \pi\delta \text{ is not empty}$
\end{itemize}
Obviously, the last choice is the most expensive and the first two should be preferred.

Whenever a new assignment is made, we first try to adjust the watched literals on level one. If a clause contains no longer two appropriate literals, we push it to the second level.
On this level we index all literals of the clauses, 
e.g.\ in a context tree with top-level symbol hashing. 

Putting clauses back to level one can be done either by maintaining an activity heuristics and time to time manually check for watchable literals, or managing lists of pointers for all clause-literals to relevant assignments 
on the trail.

This topology should make propagation cheaper, and in particular 
using level one should make it easier to ignore clauses irrelevant w.r.t.\ the recent assignments.

%% file: src/s8-tech/p2-cand.tex
%
Before we propose an abstract algorithm for exhaustive propagation, 
we introduce a simple derivation system for finding candidates. 
Of course, in the actual implementation this system will be replaced by more efficient algorithms on 
the indexing structures. 
\leaveabit\noindent
The rules work on tuples of the form $(C; \sigma; \pi)^i$ where
\begin{itemize}
	\item $C$ is a clause, a subclause of some initial clause $C_0$ from the current clause set
	\item $\sigma$ is a substitution over $\mVar(C_0)$
	\item $\pi$ is a dismatching constraint 
	\item $i$ is the number of application of the last assignment of the trail, which has relevance in the next section
\end{itemize}
The initial tuple for a clause $C_0 \in \text{N}\cup\text{U}$ is $(C_0; \emptyset; \top)^0$ and we try to resolve each literal in $C$ with the following rule:
\[
	(C \lor L; \sigma; \pi)^i \Rightarrow_{\Gamma} 
	(C; \sigma\theta; \pi\theta \land \pi'\theta)^{i'}
\]
Where there is a $(\neg L'\cdot\sigma'; \pi')^{\alpha} \in \Gamma$ such that
\begin{itemize}
	\item $\exists\theta = \mMGU(L\sigma, L'\sigma')$
	\item $(\pi\theta \land \pi'\theta) \ne \bot$ and normalized
	\item $i'$ is $i +1$ if $(\neg L'\cdot\sigma'; \pi')^{\alpha}$ is the last assignment in $\Gamma$, and $i$ otherwise
\end{itemize}
Applying this rule we can get candidates for the rules \emph{Conflict}, and \emph{Propagate} by deriving respectively tuples of the form
\begin{itemize}
	\item $(\bot; \sigma; \pi)^i$, or
	\item $(L; \sigma; \pi)^i$
\end{itemize}
We note that non-emptiness is not checked fully, only a cheaper precondition of it. 
Free left-hand side variables and already defined instances are not removed either.

%% file: src/s8-tech/p3-prop.tex
In this section, we propose the abstract algorithm \texttt{PROP} for exhaustive propagation with conflict detection.
It basically processes a queue \texttt{PQ} of candidates for new assignments. 
As an invariant, we assume each constrained literal in the queue 
\begin{enumerate}
	\item has a normalized non-$\bot$ constraint
	\item consistent with the current $\Gamma$
	\item contains no free left-hand side variable
\end{enumerate}
\subsubsection*{PROP}
Initially, this queue consists of the literals induced by the unit clauses. 
Unit clauses has to be checked for contradiction prior calling \texttt{PROP}.
When calling after decisions, \texttt{PQ} is assumed to contain the 
immediate consequences of the decision. 
Checking for blocking should generate this set anyway.

\texttt{PROP} processes the literals on \texttt{PQ}.
First, it removes already defined instances by calling the function \texttt{DIFF}.
This produces a set of disjoint and undefined constrained literals, each of which
is a valid subject of \emph{Propagate}.
See Section 2.3 for the definition of the difference operation "$-$", and see below 
the abstract algorithm for \texttt{DIFF}.

These literals are then checked for emptiness, added to $\Gamma$ and set to true.
Their consequences - conflicts and new candidates for \texttt{PQ} - 
are then generated by \texttt{addConsequences}.

We continue this process until \texttt{PQ} gets empty, or a conflict is found.
The first indicates the finished exhaustive application of \emph{Propagate}, and \emph{Decide} can be called. 
In this case we return $true$. And in the latter case, we return $false$, and the found conflict is 
stored in \texttt{conflictSet}.

On the course of this section, we might use the symbol $\ell$ to denote annotated constrained literals, and the following 
auxiliary functions:
\begin{itemize}
	\item \texttt{pop}: removes an element of a queue, list, or set
	\item \texttt{notEmpty}: carries out a full non-emptiness check for a constraint or constrained literal
	\item \texttt{addAssignment}: adds a new assignment to $\Gamma$ (and its indexing structures)
	\item \texttt{cUNIF($\ell$, $\Gamma$)}: Finds the literals in $\Gamma$ which are unifiable with $\ell$, and returns an array of them and its size
	\item \texttt{NF}: normalizes a constraint, constrained literal, or a set of constrained literals, as described in Subsection 2.2. In the latter case, it removes resulting literals with $\bot$-constraints.
	\item \texttt{freeLVars}: produces the set of free left-hand side variables of a constrained literal
	\item \texttt{selectOne}: randomly, or heuristically selects an element of a set, or a list
	\item \texttt{adjustLevel1}: adjusts the first index level for clauses after a new assignment given as parameter, as described in Section 6.2.
	\item \texttt{getCandidates}: provides the list of indexed clauses which contains a literal unifiable with the complement of a given literal
\end{itemize}
%
\input{src/s8-tech/alg-prop}
\subsubsection*{DIFF}
It iteratively removes the already defined instances from the proposed assignment. 
The result is a set of disjoint and undefined constrained literals with non-$\bot$ constraints.
\input{src/s8-tech/alg-diff}
\subsubsection*{elimFV}
An auxiliary function for finding new candidates. 
It iteratively removes the free left-hand side variables, and only keeps the 
literals with non-$\bot$ normalized constraints.
\input{src/s8-tech/alg-elimFV}
\subsubsection*{addConsequences}
Finally, \texttt{addConsequences} checks whether a new assignment produces a conflict and generates new 
candidates for \texttt{PQ}. It returns $true$ if no conflict is found, and $false$ otherwise. 
If a conflict is found, it is saved in \texttt{conflictSet}. 

We distinguish two types of conflicts.
It is easy to see, that if the new assignment is used only once in deriving a conflict, then 
\texttt{PQ} must already hold an unprocessed candidate which is falsified by the new assignment.
Thus, we check \texttt{PQ} first for a contradiction, and start generating new candidates with $\Rightarrow_{\Gamma}$
only afterwards.

We then use the derivation system of 6.3 to derive new constrained literals. We only consider derivations 
where the latest assignment has to be used at least once. If it is used only once we can be sure the new 
literal is not false. If it is not the case, we check for a possible conflict.

As stated before, in the actual implementation the proper retrieval algorithms will eliminate the inefficiency of considering all derivations.

Finally, the new candidates are tested for free variables, and they are removed if there are any.
\input{src/s8-tech/alg-addCons}

%% file: src/s8-tech/alg-prop.tex
\begin{function}
	\SetKwData{PQ}{PQ}
  \SetKwData{N}{N}\SetKwData{U}{U}
	\SetKwFunction{pop}{pop}
	\SetKwFunction{DIFF}{DIFF}
	\SetKwFunction{notEmpty}{notEmpty}
	\SetKwFunction{Assign}{addAssignment}
	\SetKwFunction{Conseq}{addConsequences}
	
	\caption{PROP(N, U, $\Gamma$, PQ)}
	\While{\PQ $\ne$ $\emptyset$}{
		$\ell = (L\cdot\sigma; \pi)^C$ $\gets$ \pop{\PQ}\;
		$\Delta$ $\gets$ \DIFF{$\ell$, $\Gamma$}\;
		\ForEach{$\ell' = (L'\cdot\sigma'; \pi')^C \in \Delta$}{
			\If{\notEmpty{$\ell'$}}{
				\tcp{Applying Propagate:}
				\Assign{$\Gamma$, $\ell'$}\;
				\lIf{\Conseq{\N,\U,$\Gamma$,$\ell'$,\PQ} = false}{\KwRet{false}}
			}
		}
	}
	\KwRet{true}\;
\end{function}

%% file: src/s8-tech/alg-diff.tex
\begin{function}
	\SetKwFunction{UNIF}{cUNIF}
	\SetKwFunction{NF}{NF}
	\SetKwData{k}{k}\SetKwData{i}{i}
	
	\caption{DIFF($\ell^*$, $\Gamma$)}
	($\vec{\ell}$, \k) $\gets$ \UNIF{$\ell^*$, $\Gamma$}\;
	$\Delta_0$ $\gets$ $\{\ell^*\}$\;
	\For{\i = $1$, \dots, \k}{
		$\Delta_i$ $\gets$ $\operatorname{NF}(\Delta_{i-1} - \vec{\ell}[i])$\;
	}	
	\KwRet{$\Delta_k$}\;
\end{function}

%% file: src/s8-tech/alg-elimFV.tex
\begin{function}
	\SetKwFunction{pop}{pop}
	\SetKwFunction{FV}{freeLVars}
	\SetKwFunction{NF}{NF}
	\SetKwFunction{sel}{selectOne}
	
	\caption{elimFV($\ell_0$)}
	\tcp{Prereq: $\operatorname{NF}(\pi_{\ell_0}) \ne \bot$}
	$\Delta$, $\Delta^*$ $\gets$ $\{\ell_0\}$, $\emptyset$\;
	\While{ $\Delta \ne \emptyset$ }{
		$\ell = (L\cdot\sigma; \pi)^C$ $\gets$ \pop{$\Delta$}\;
		\If{\FV{$\ell$} = $\emptyset$}{
			$\Delta^*$ $\gets$ $\Delta^* \cup \{\ell\}$\;
		}
		\Else{
			$x$ $\gets$ \sel{\FV{$\ell$}}\;
			\tcp{Instantiation with each constant from the domain:}
			\ForEach{$d \in \mDomain$}{
				$\pi'$ $\gets$ \NF{$\pi\{x \gets d\}$}\;
				\lIf{$\pi' \ne \bot$} {$\Delta$ $\gets$ $\Delta \cup \{ (L\cdot\sigma\{x \gets d\}; \pi')^C \}$}
			}
		}
	}
	\KwRet{$\Delta^*$}
\end{function}

%% file: src/s8-tech/alg-addCons.tex
\begin{function}
	\SetKwData{PQ}{PQ}
	\SetKwData{CS}{conflictSet}
	\SetKwFunction{notEmpty}{notEmpty}
	\SetKwFunction{NF}{NF}
	\SetKwFunction{FV}{freeLVars}
	\SetKwFunction{elimFV}{elimFV}
	\SetKwFunction{getCandidates}{getCandidates}
	\SetKwFunction{adjLvl}{adjustLevel1}
	
	\caption{addConsequences(N,U,$\Gamma$,$\ell'$, PQ)}
	$(L'\cdot\sigma'; \pi')^\alpha$ $\gets$ $\ell'$\;
	\tcp{Step 1: Check Type-1 Conflicts}
	\ForEach{$\ell = (L\cdot\sigma; \pi)^C$ $\in$ \PQ}{
		\If{$\exists\delta = \mMGU(\neg L\sigma, L'\sigma')$ and \NF{$\pi\delta\land\pi'\delta$}$\ne \bot$ and \notEmpty{$\pi\delta\land\pi'\delta$}}{
			\CS $\gets$ $(C; \sigma\delta; \pi\delta\land\pi'\delta)$\;
			\KwRet{false}\;
		}
	}
	\tcp{Step 2: Adjust indexing level 1}
	\adjLvl{N,U,$\ell'$}\;
	\tcp{Step 3: Generate consequences}
	\ForEach{$C$ $\in$ \getCandidates{$\text{N}\cup\text{U}$, $\ell'$}}{
		\ForEach{derivation $(C; \emptyset; \top)^0 \Rightarrow_\Gamma^* (L^*; \sigma^*; \pi^*)^i$ with $i \geq 1$}{
			\tcp{Check Type-2 Conflicts}
			\If{$i \geq 2$}{
				\If{$\exists \ell = (L\cdot\sigma; \pi)^\beta \in \Gamma$ such that
								$\exists \delta = \mMGU(\neg L\sigma, L^*\sigma^*)$ and
								\NF{$\pi\delta\land\pi^*\delta$} $\ne \bot$ and
								\notEmpty{$\pi\delta\land\pi^*\delta$}
						}
				{
					\CS $\gets$ $(C; \sigma^*\delta; \pi\delta\land\pi^*\delta)$\;
					\KwRet{false}\;
				}
			}
			$\ell^*$ $\gets$ $(L^*\cdot\sigma^*; \pi^*)^C$\;
			\lIf{\FV{$\ell^*$} = $\emptyset$}{\PQ $\gets$ \PQ, $\ell^*$}
			\lElse{\PQ $\gets$ \PQ, \elimFV{$\ell^*$}}
		}
	}
	\KwRet{true}\;
\end{function}
%

%% file: src/s8-tech/ss2a-decisions.tex
%
When making a new decision, we a pick a candidate $(L; \pi)$, 
remove all the already defined instances, and then 
test all immediate conflicts for blocking. 

If there is a blocking conflict, 
we might then either pick an entirely new decision candidate, or try to \emph{fix} $(L;\pi)$ by instantiating some variables in $L$, and thereby generating a new set of candidates.  

This can be achieved by picking a blocking ground instance $C' = C\delta$ which contains $\neg L\delta_1, \neg L\delta_2$ such that
both $L\delta_1, L\delta_2 \in \mGnd(L; \pi)$ and $L\delta_1 \ne L\delta_2$ holds. 
Now, choose a variable for which $x\delta_1 \ne x\delta_2$, and
split $(L; \pi)$ into $(L\{x \gets x\delta_1\}; \pi\{x \gets x\delta_1\})$ and $(L; \pi \land x \ne x\delta_1)$.
By instantiating further variables, we eventually get a decision which is not blocking, since a ground decision is always suitable.

A non-blocking decision is then added to $\Gamma$, and whether we found a non-blocking conflict or not, we continue with conflict 
resolution or with calling \texttt{PROP} after generating the immediate propagation candidates in a similar way as in \texttt{addConsequences}.

Initially, the set of decision candidates are generated from the literals occurring in $\text{N}$. 
This set can be later refined by the above steps, and individual candidates might be substituted with sets of new candidates.

Since removing defined instances is always relative to the current $\Gamma$, it has to be guaranteed that the set of all possible candidates
covers the original set.
It can be ensured for example by keeping a trail for these refinement steps as well, and re-roll them in parallel with the backtracking procedure.
%

%% file: src/s8-tech/ss2-vsids.tex
%
Most current SAT solvers also employ variable selection schemes based on dynamic ranking of propositional variables. 
This technique rewards variables involved in recent conflicts, and proved itself efficient in the propositional context. 

Following the footsteps of the now classic \emph{decaying variable sum}, we reward the literals 
involved in the clause learning phase following the latest conflict. 

This is accomplished by maintaining a list of literals and scores. 
Whenever some literal $L$ is added to the clause of the intermediate state, we add a pair $(L; v)$ to this list.

To focus on recent conflicts, we increase $v$ gradually, and occasionally we reset $v$ to some initial value and normalize the list.
The latter can be triggered upon reaching some extreme value, automatically after a certain number conflicts, or 
at restarts. 

Restarts are commonly used in SAT solvers to redirect the focus of the search using the learned clauses and the current variable scores. 
Applying it only finitely many times does not violate completeness.

Then, whenever we need to choose a new decision, we rank the candidates by combining the scores belonging to literals which are unifiable with
the candidate in question. As an example we propose addition or maximum. We then choose the literal with the highest combined score.
%

%% file: src/s8-tech/ss3-cl.tex
%
As the conflicts are now discovered, every conflict-set uniquely assigns a $\Gamma$-assignment to each literal of 
the conflict clause. This make detecting assertiveness easy and spares us a number of emptiness checks, as they 
are already done during conflict detection. 
This way, the only non-deterministic choice is the application of \emph{Factorize} versus \emph{Resolve}, when both is applicable. 

Once a new clause is learned, a suitable backtrack level is needed. 
Should we learn only the ground clauses in $\mGnd(C\sigma; \pi)$ when the last conflict-set  is $(C; \sigma; \pi)$, 
we could determine the backtrack position at ease, similarly to the propositional solvers. 

But we learn the more general $C$, and the right backtrack position has to be computed from all the 
instances of $C$.
We have to consider all conflicting instance of $C$ w.r.t.\ $\Gamma$, and for each instance, we have to determine a minimal backtrack position.
Then, we backjump to the minimum of these positions. 

Without providing more details, we only note that some instances produce new assignments after backtrack, some might block existing decisions, and some might even
be new conflicts after backjump.

%% file: src/s6-relw/related-work.tex
%
\section{Related Work}
In this section, we briefly compare NRCL to existing solutions.
As \mEPR~problems can be successfully handled with finite model finders as well, 
we cover both \mEPRs-specific techniques and more general finite model building approaches.
In the case of the latter systems, we focus on their behavior on the \mEPR~fragment.

The first successful approaches to finite model building were \emph{Mace} and \emph{SEM}, see e.g.\ \cite{Tammet03finitemodel}.
The early version of \emph{Mace} flattens and grounds the given clause set, and passes it on to a CDCL-based SAT solver. 
This approach is developed further by \emph{Paradox}~\cite{ParadoxRef}. 

Compared to this approach, we work directly with the first-order clause set instead of the often exponentially larger set of ground instances.

The latest version of \emph{Mace}~\cite{Mace4Ref03} follows the approach of \emph{SEM}~\cite{SEMRef95} and \emph{FINDER}~\cite{FinderRef94}. 
Instead of generating the ground instances, 
it maintains the function and predicate tables, and fills them out using a sophisticated backtracking algorithm. 

Compared to this approach, we represent the model implicitly via constrained literals,  
and let the learned clauses guide our calculus.

Over the last decade several attempts were made to lift CDCL and its ancestor, DPLL - 
a calculus using backtracking instead of backjumping and clause learning. 
\emph{Model Evolution}~\cite{ModelEvolution03} and its implementation \emph{Darwin}~\cite{Darwin04}
represents a model with a set of first-order literals, called \emph{context}, and 
detects conflicts using syntactic concepts weaker then the full-fledged semantics based on
the induced interpretation.
This potentially leads to longer derivation before detecting a false clause.

It is refutationally complete over first-order clauses and provides a 
decision procedure for the {\mEPR} fragment. 
Its extension~\cite{ModelEvolutionLemma06} enriches the calculus with learning lemmas at conflicts, 
and uses backjumping instead of the original backtracking approach. 

Compared to \emph{Model Evolution}, NRCL relies on the full-fledged semantics,  
and we learn only non-redundant clauses.
It is not clear if the latter holds for \emph{Model Evolution}, 
especially the admissibility of the classic criterions
needs in-depth considerations.

Finally, it was shown in~\cite{FermullerCADE05} that using contexts might result in exponentially larger model representations.
We note that this result holds for the general case with function symbols, but in our setting \eg the constrained literal
\[(P(x_1, x_2, \dots, x_k); x_1 \ne x_2 \land x_2 \ne x_3 \land \dots \land x_{k-1} \ne x_k)\]
whose size is $O(k)$, requires a representation of size at least $O(k^2)$ as a context.
Thus, at least a quadratic relation holds even for the {\mEPR} fragment.

\emph{DPLL(SX)}~\cite{DPLLSX10} attempts to lift CDCL to {\mEPRs} in the same manner as we do, 
has an almost identical rule set, 
and uses substitution sets represented by BDDs as constraints. 
Substitution sets provide an explicit way to represent models. 

It is well-known that in the general setting with function symbols implicit representations have stronger expressive power~\cite{Pichler03}\cite{LassezM87}.
In our setting, explicit representations have the potential to be exponentially larger 
then the corresponding implicit representations.

The following simple example demonstrates this claim.
Over $\mDomain = \{a,b,c\}$, consider the constrained literal
\[(P(x_1, x_2, \dots, x_k); x_1 \ne x_2 \land x_2 \ne x_3 \land \dots \land x_{k-1} \ne x_k)\]
Then it is easy to see that the corresponding explicit representation is made up of all the ground instances covered by this literal.

Therefore, while the size of the implicit representation increases linearly in $k$, the size of the corresponding explicit representation
is $O(2^k)$, \ie increases exponentially in $k$.

The authors of this paper are convinced that this exponential blow-up happens whenever in the implicit representation has 
no finite explicit representation (see \cite{Pichler03}\cite{LassezM87} for details) in the language enriched with a function symbol. 
However, this conjuncture needs further consideration, and we leave it for future work.

Furthermore, compared to \emph{DPLL(SX)} our approach is more modular 
as it allows the use of an arbitrary constraint language, restricted only by the operations we expect. 
Dismatching constraints can be extended beyond the {\mEPR} fragment easily, while in 
the case of BDD-encodings, it is not trivial.

DPLL(SX) also lacks the concept for blocking, and applies an explicit refine rule instead.
As a side effect, it learns nothing from conflicts which lead to blocking clauses, and 
in these cases it abandons conflict resolution and refines the last decision.
Finally, we also address redundancy, and exploit the non-redundancy result to show termination, which we consider a valuable addition.

The most recent calculus \emph{SGGS}, introduced in~\cite{SGGSExposition}, promises a semantically guided, goal sensitive, model-based proof system. 
It uses simple constraints, so-called \emph{standard forms}, conjunctions of negative atomic constraints 
of the form $x \ne y$, or $\operatorname{top}(x) \ne f$.

Then, a model is represented by a sequence of constrained clauses with selected literals. 
This sequence overrides a given initial interpretation $I$, 
which serves both as initial model assumption and as semantic guidance for the calculus.

The procedure then keeps expanding this sequence in order to satisfy more and more clauses, and handles contradictions via resolution and splitting 
the constrained clauses to maintain an invariant - 
every literal in every clause in the sequence must have \emph{either only false, or only true instances} \wrt 
$I$ and the constraints.

NRCL utilizes a more expressive constraint language, which allows tuples to be used.
This results in less fragmentation of the representation, \ie \emph{SGGS} might need several constraints in standard form
to express a single dismatching constraint of our calculus. 

This allows us to learn more general clauses, and also potentially decreases the size of the representation. 
Our model representation relies on constrained literals instead of clauses, and we consider it to be more
explicit than the approach of \emph{SGGS} which requires identifying the constrained instances of the clauses which are indeed
producing new assignments.

Finally, the resolution applied by \emph{SGGS} only repairs the model, it can be discarded later as the search progresses, and the splittings applied to maintain the invariants 
also forces the result of resolution to be more specific, more local. 
Compared to this, our calculus learns and saves new clauses, uses backjumping, and we proved these clauses are non-redundant.

We also mention \emph{geometric resolution}~\cite{GeoRes06} which uses a special normal form called \emph{geometric normal form}. In this calculus the formulas
themselves constitute the rules of a system based on backtracking. 
Through the inference \emph{geometric resolution} it also provides a way to learn new formulas.
The transformation to geometric normal form also includes flattening, 
which our approach avoids.

The calculus \emph{Inst-Gen}~\cite{InstGen03} and its implementation \emph{iProver}~\cite{iProver08} 
has been quite successful at solving {\mEPR} problems, and competitive even for the first-order fragment.
It generates a propositional approximation of the clause set by instantiating all the variables with constants, 
and passes it on to a CDCL-based SAT solver. 

Unsatisfiability of the approximation entails the unsatisfiability of the original problem.
On the other hand, if an abstract model is generated, it is used to guide the calculus to add 
proper instances of the original clauses, which refines the propositional abstraction.

This procedure is continued then, until either unsatisfiability is proven, or saturation is achieved, 
which implies that the abstract model can be lifted to a first-order model for the original clause set.

The algorithm is further enhanced by using dismatching constraints, and applying redundancy elimination based 
on generating first-order resolvents for subsumption with a theorem prover, and finding simplification candidates efficiently
with ground reasoning.

Compared to \emph{iProver}, our approach is fine-grained, 
as the evaluation and refinement of our abstraction happen interleaved with the other reasoning steps.
Furthermore, we work directly with the original clause set, and our trail always corresponds 
to a consistent first-order model candidate.

In addition to the theoretical comparison, we also ran a small experiment for models represented by literals of the form 
$(P(x_1,\ldots,x_k);x_1\neq x_2,\ldots,x_{k-1}\neq x_k)$
The clause set

  $Q(x,x), \neg Q(a_1, a_2),  \neg Q(a_2, a_3), \ldots, \neg Q(a_{n-1}, a_n),$
	
  $\neg P(x_1,x_1, x_3,\ldots, x_k), \ldots, \neg P(x_1,x_2,\ldots, x_{k-1},x_{k-1}),$
	
  $\neg Q(x,z)\lor Q(x,y)\lor Q(y,z), P(x_1,\ldots,x_k) \vee Q(x_1,x_2)\vee\ldots\vee Q(x_{k-1}, x_k)$\newline
has a model where the positive atoms are represented by the constrained literals 
$(P(x_1,\ldots,x_k);x_1\neq x_2,\ldots,x_{k-1}\neq x_k)$
and $(Q(x,x);\top)$. NRCL directly finds this model, \ie without backjumping even once, 
by exhaustively applying propagation, making a single decision on $P(x_1,\ldots,x_k)$ and finally 
setting all undefined $Q(x,y)$ literals to false. Furthermore, any regular run would find a similar model without 
backjumping even once.

We tested this clause set with the available state-of-the-art provers \emph{Darwin}
(1.4.5) and \emph{iProver} (0.8.1).
The experiments were carried out on a Debian Linux
(4.7.2-5) Intel (Xeon E5-2680, 2.7GHZ) computer with 256GB physical memory.
For $n=7$ and $k=5,7,9$, \emph{Darwin} needs $0.2, 8.1, 518$ seconds to find a model, respectively.
For $k=7$ and $n=7,10,13$, \emph{Darwin} needs $8.1, 62, 347$ seconds to find a model, respectively.
For $k=9$ and $n=7,10,13,16,19,22$, \emph{iProver} needs $0, 0.2, 21, 39, 116, 718$ seconds to find a model, 
respectively.

In the case of \emph{Darwin}, these results show an exponentially 
growing solution time \wrt $k$ (the arity of $P$) or  $n$ (the domain size). 
\emph{iProver} is robust against increasing $k$ but not against increasing
$n$, where it also shows an exponential growth. 
 This shows that our model representation is not subsumed by either \emph{Darwin} or \emph{iProver}.

%

Finally, even general purpose first-order theorem provers implement specialized techniques to handle {\mEPR} problems.

\emph{Generalisation} introduced in~\cite{Generalization08} for \emph{Vampire} 
is an additional technique for resolution-based saturation. 
It infers $P(x)$ if $P(c)$ has been established for all relevant constant $c$.
Coupled with efficient sort inference, it has the potential to exponentially speed up theorem proving.

The technique introduced in~\cite{HillenbrandWeidenbach13} for \emph{SPASS} employs 
a combination of restricted superposition on Horn clauses, and \emph{labelled splitting}~\cite{FietzkeW09} on non-Horn clauses.

Compared to these approaches, NRCL maintains a model candidate, it is restricted to learn clauses only at conflicts and only non-redundant ones, 
does not rely on Horn clauses, and the implicit branchings through decisions and backjumps are more elaborate and guided by the model search,
compared to the splitting techniques employed by first-order theorem provers.
However, we note that for some problem classes finite superposition saturation is still
superior to explicit model generation, see \eg superposition for knowledge bases in~\cite{SudaWeidenbachWischnewskiIJCAR10}.
%

%% file: src/s7-futw/future-work.tex
\section{Conclusion}
In this paper, we proposed the decision procedure NRCL for the {\mEPR}
fragment. 
Our approach represents a model candidate as a set of constrained literals, and 
derives a model or a proof of unsatisfiability through a series of decisions, 
propagations, and learning new clauses. 

Our work closely relates to \emph{DPLL(SX)}~\cite{DPLLSX10}, which introduces a similar calculus,
and the more recent calculus \emph{SGGS}~\cite{SGGSExposition}.
Compared to earlier work in this direction, we investigated the standard redundancy notion
w.r.t.\ the ordering induced by the current trail.

One of the main contributions of NRCL over existing work is that, by design, 
we can prove our learned clauses to be non-redundant, i.e., any learned clause makes progress 
towards finding a model or a refutation, because it eliminates at least one potential model. 
In general, we consider this a key property for automated reasoning calculi.

Projecting NRCL down to propositional logic proves this property for CDCL 
with respect to our notion of redundancy. 
Our notion also admits techniques like subsumption and subsumption resolution, 
which are important in both SAT solving and first-order theorem proving. 
We see this as a strong indication that a future implementation 
will also contribute to the state of the art.

In Section 8, we addressed some of the difficulties of this approach, and provided 
details for implementation.
Finally, we gave a brief comparison to the existing solutions in Section 9.

As future research, the immediate goal is to 
make an efficient implementation of NRCL.
This includes developing suitable and efficient term indexing structures, possibly revising the constraint language, and 
defining concrete and efficient heuristics for selecting decisions.

On the other hand, the long-term goal of our research is to extend this calculus beyond {\mEPR}.
The next step into this direction is to enrich our calculus with function symbols and sorts to handle the
\emph{non-cyclic fragment} introduced in~\cite{Noncyclic13}. 
This class still has the finite Herbrand model property, thus, our results will directly extend to this fragment.

The further goals are to consider other decidable fragments, to introduce equality into our calculus, and finally 
to extend our work to finite model finding.
%

%% file: notes/notes-main.tex
\section{Notes and Design Decisions}
\input{notes/weak-reg}

%% file: notes/weak-reg.tex
\subsection{Weak Regularity}
We note that the weaker definition for regularity, see below, is just as effective as the one we choose.
\begin{defi}
A \emph{run} is called \emph{weak-regular} iff the following holds:
	\begin{itemize}
		\item During conflict search, rules are always applied in this order exhaustively: terminal rules, \emph{Conflict}, \emph{Propagate}, \emph{Decide}.
		\item In conflict resolution \emph{Backjump} is always applied as soon as possible.
	\end{itemize}
\end{defi}
Some properties however won't hold, in particular the claims (1) and (2) in Lemma~\ref{lemmRegProps} no longer hold, as demonstrated 
by the example below.
\begin{example}
Let $\mDomain = \{ a, b, c, d, e, f, l, k \}$, and $\text{N}$ the following clause set:
\begin{center}\begin{tabular}{rlll}
$\{$ & $\exClauseNo{0}:~\neg P(a), $ & $\exClauseNo{1}:~P(a) \lor \neg P(d) \lor \neg P(l), $ & \\
       & $\exClauseNo{2}:~P(l) \lor P(k),$ & $\exClauseNo{3}:~P(a) \lor P(l) \lor \neg P (k), $ & \\
       & $\exClauseNo{4}:~\neg P(b) \lor P(d) \lor \neg P(l), $ & $\exClauseNo{5}:~\neg P(c) \lor \neg P(e) \lor \neg P(f)$ & $\}$ \\
\end{tabular}\end{center}
Then, the next weak-regular derivation demonstrates both learning the same clause $P(a) \lor P(l)$ twice, 
and deducing the literal $(P(l); \top)^{P(a) \lor P(l)}$ 
contradicting the stronger regularity notion of Definition~\ref{regRunDef}.
\[
(\epsilon; \text{N}; \emptyset; 0; \top)
 \stackrel{*}{\Rightarrow}
(\neg P(a)^{\exClauseNo{0}}, P(b)^1, \neg P(c)^2, P(d)^3, \neg P(l)^{\exClauseNo{1}}, P(k)^{\exClauseNo{2}}; \text{N}; \emptyset; 3; \top)
\]
\[
\stackrel{Conflict(\exClauseNo{3})}{\Rightarrow}
(\neg P(a)^{\exClauseNo{0}}, P(b)^1, \neg P(c)^2, P(d)^3, \neg P(l)^{\exClauseNo{1}}, P(k)^{\exClauseNo{2}}; \text{N}; \emptyset; 3; (\exClauseNo{3};\emptyset; \top))
\]
\[
\stackrel{ConfRes*}{\Rightarrow}
(\neg P(a)^{\exClauseNo{0}}, P(b)^1, \neg P(c)^2, P(d)^3, \neg P(l)^{\exClauseNo{1}}; \text{N}; \emptyset; 3; (P(a) \lor P(l);\emptyset; \top))
\]
\[
\stackrel{Backjump}{\Rightarrow}
(\neg P(a)^{\exClauseNo{0}}, P(b)^1, \neg P(c)^2; \text{N}; \{\exClauseNo{6}:~P(a) \lor P(l)\}; 2; \top)
\]
Let $\text{U}_1 = \{\exClauseNo{6}:~P(a) \lor P(l)\}$.
\[
\stackrel{*}{\Rightarrow}
(\neg P(a)^{\exClauseNo{0}}, P(b)^1, \neg P(c)^2, P(l)^{\exClauseNo{6}}, \neg P(d)^{\exClauseNo{1}}; \text{N}; \text{U}_1; 2; \top)
\]
\[
\stackrel{Conflict(\exClauseNo{4})}{\Rightarrow}
(\neg P(a)^{\exClauseNo{0}}, P(b)^1, \neg P(c)^2, P(l)^{\exClauseNo{6}}, \neg P(d)^{\exClauseNo{1}}; \text{N}; \text{U}_1; 2;(\exClauseNo{4};\emptyset;\top))
\]
\[
\stackrel{ConfRes*}{\Rightarrow}
(\neg P(a)^{\exClauseNo{0}}, P(b)^1, \neg P(c)^2, P(l)^{\exClauseNo{6}}; \text{N}; \text{U}_1; 2;(P(a) \lor \neg P(b) \lor \neg P(l);\emptyset;\top))
\]
\[
\stackrel{Backjump}{\Rightarrow}
(\neg P(a)^{\exClauseNo{0}}, P(b)^1; \text{N}; \text{U}_1 \cup \{\exClauseNo{7}:~P(a) \lor \neg P(b) \lor \neg P(l)\}; 1;\top)
\]
Let $\text{U}_2 =  \text{U}_1 \cup \{\exClauseNo{7}:~P(a) \lor \neg P(b) \lor \neg P(l)\}$.
\[
\stackrel{Propagate}{\Rightarrow}
(\neg P(a)^{\exClauseNo{0}}, P(b)^1, \neg P(l)^{\exClauseNo{7}}; \text{N}; \text{U}_2; 1; \top)
\]
\[
\stackrel{Conflict(\exClauseNo{6})}{\Rightarrow}
(\neg P(a)^{\exClauseNo{0}}, P(b)^1, \neg P(l)^{\exClauseNo{7}}; \text{N}; \text{U}_2; 1; (\exClauseNo{6}; \emptyset; \top))
\]
And, since $\exClauseNo{6}$ is already assertive at this point, we learn it again and backjump.
\[
\stackrel{Backjump}{\Rightarrow}
(\neg P(a)^{\exClauseNo{0}}; \text{N}; \text{U}_2 \cup \{\exClauseNo{8}: P(a) \lor P(l)\}; 0; \top)
\]
Above, for any literal $L$ and annotation $\alpha$, $L^\alpha$ is a short-hand for $(L; \top)^\alpha$.
\end{example}
Soundness and termination could still be proved, and we believe non-redundant learning would also hold
with the exception of immediate conflicts of the kind demonstrated above, i.e. when after an inaccurate 
backjump we end up with an assertive conflict clause and learn the clause itself again.

We could modify \emph{Backjump} to identify this situation and make a proper backjump when we 
would learn an already known clause again.
We believe however that Lemma~\ref{lemmRegProps} is a useful invariant and easy to take granted, thus, we 
decided to use a stronger notion of regular runs in Definition~\ref{regRunDef}.

%% file: alagi_NRCLfull.bbl
\begin{thebibliography}{10}

\bibitem{DBLP:conf/cade/2008}
A.~Armando, P.~Baumgartner, and G.~Dowek, eds.
\newblock {\em Automated Reasoning, 4th International Joint Conference, IJCAR
  2008, Sydney, Australia, August 12-15, 2008, Proceedings}, 2008, {\em LNCS
  5195}. Springer.

\bibitem{BachmairG01}
L.~Bachmair and H.~Ganzinger.
\newblock Resolution Theorem Proving.
\newblock In Robinson and Voronkov \cite{DBLP:books/el/RobinsonV01}, pp.
  19--99.

\bibitem{BachmairGLS92}
L.~Bachmair, H.~Ganzinger, C.~Lynch, and W.~Snyder.
\newblock Basic Paramodulation and Superposition.
\newblock In D.~Kapur, ed., {\em Automated Deduction - CADE-11, 11th
  International Conference on Automated Deduction, Saratoga Springs, NY, USA,
  June 15-18, 1992, Proceedings}, 1992, {\em LNCS 607}, pp. 462--476. Springer.

\bibitem{BachmairGLS95}
L.~Bachmair, H.~Ganzinger, C.~Lynch, and W.~Snyder.
\newblock Basic Paramodulation.
\newblock {\em Inf. Comput.}, 121(2):172--192, 1995.

\bibitem{Darwin04}
P.~Baumgartner, A.~Fuchs, and C.~Tinelli.
\newblock Darwin: A Theorem Prover for the Model Evolution Calculus.
\newblock In S.~Schulz, T.~Tammet, and G.~Sutcliffe, eds., {\em Proceedings of
  the 1st Workshop on Empirically Successful First Order Reasoning (ESFOR'04)},
  2004, IJCAR 2004 Workshop Proceedings, pp. 1--24. UCC.

\bibitem{ModelEvolutionLemma06}
P.~Baumgartner, A.~Fuchs, and C.~Tinelli.
\newblock Lemma Learning in the Model Evolution Calculus.
\newblock In M.~Hermann and A.~Voronkov, eds., {\em LPAR}, 2006, {\em LNCS
  4246}, pp. 572--586. Springer.

\bibitem{ModelEvolution03}
P.~Baumgartner and C.~Tinelli.
\newblock The Model Evolution Calculus.
\newblock In F.~Baader, ed., {\em CADE}, 2003, {\em LNCS 2741}, pp. 350--364.
  Springer.

\bibitem{SATHandbook}
A.~Biere, M.~Heule, H.~van Maaren, and T.~Walsh, eds.
\newblock {\em Handbook of Satisfiability}, 2009, {\em Frontiers in Artificial
  Intelligence and Applications}, vol. 185. IOS Press.

\bibitem{SGGSExposition}
M.~P. Bonacina and D.~A. Plaisted.
\newblock SGGS theorem proving: an exposition.
\newblock {\em Notes of the Fourth Workshop on Practical Aspects in Automated
  Reasoning (PAAR), Seventh International Joint Conference on Automated
  Reasoning (IJCAR) and Sixth Federated Logic Conference (FLoC), Vienna,
  Austria, July 2014.}, 2014.

\bibitem{ParadoxRef}
K.~Claessen and N.~Sörensson.
\newblock New Techniques that Improve MACE-style Finite Model Finding.
\newblock In {\em Proceedings of the CADE-19 Workshop: Model Computation -
  Principles, Algorithms, Applications}, 2003.

\bibitem{Comon91}
H.~Comon.
\newblock Disunification: A Survey.
\newblock In {\em Computational Logic - Essays in Honor of Alan Robinson},
  1991, pp. 322--359.

\bibitem{EiterFT05}
T.~Eiter, W.~Faber, and P.~Traxler.
\newblock Testing Strong Equivalence of Datalog Programs - Implementation and
  Examples.
\newblock In C.~Baral, G.~Greco, N.~Leone, and G.~Terracina, eds., {\em Logic
  Programming and Nonmonotonic Reasoning, 8th International Conference, {LPNMR}
  2005, Diamante, Italy, September 5-8, 2005, Proceedings}, 2005, {\em LNCS
  3662}, pp. 437--441. Springer.

\bibitem{EmmerKKV10}
M.~Emmer, Z.~Khasidashvili, K.~Korovin, and A.~Voronkov.
\newblock Encoding industrial hardware verification problems into effectively
  propositional logic.
\newblock In R.~Bloem and N.~Sharygina, eds., {\em Proceedings of 10th
  International Conference on Formal Methods in Computer-Aided Design, {FMCAD}
  2010, Lugano, Switzerland, October 20-23}, 2010, pp. 137--144. {IEEE}.

\bibitem{FermullerCADE05}
C.~G. Ferm{\"u}ller and R.~Pichler.
\newblock Model Representation via Contexts and Implicit Generalizations.
\newblock In R.~Nieuwenhuis, ed., {\em CADE}, 2005, {\em LNCS 3632}, pp.
  409--423. Springer.

\bibitem{FietzkeW09}
A.~Fietzke and C.~Weidenbach.
\newblock Labelled splitting.
\newblock In {\em Ann. Math. Artif. Intell. Vol. 55 No. 1-2}, 2009, pp. 3--34.

\bibitem{InstGen03}
H.~Ganzinger and K.~Korovin.
\newblock New Directions in Instantiation-Based Theorem Proving.
\newblock In {\em LICS}, 2003, pp. 55--64. IEEE Computer Society.

\bibitem{Higman52}
G.~Higman.
\newblock {Ordering by Divisibility in Abstract Algebras}.
\newblock {\em Proceedings of the London Mathematical Society},
  s3-2(1):326--336, 1952.

\bibitem{HillenbrandWeidenbach13}
T.~Hillenbrand and C.~Weidenbach.
\newblock Superposition for Bounded Domains.
\newblock In M.~P. Bonacina and M.~Stickel, eds., {\em McCune Festschrift},
  2013, {\em LNCS 7788}, pp. 68--100. Springer.
\newblock Based on the Research Report MPI-I-2007-RG1-002.

\bibitem{HustadtMS04}
U.~Hustadt, B.~Motik, and U.~Sattler.
\newblock Reducing SHIQ-Description Logic to Disjunctive Datalog Programs.
\newblock In D.~Dubois, C.~A. Welty, and M.~Williams, eds., {\em Principles of
  Knowledge Representation and Reasoning: Proceedings of the Ninth
  International Conference (KR2004), Whistler, Canada, June 2-5, 2004}, 2004,
  pp. 152--162. {AAAI} Press.

\bibitem{KhasidashviliKV09}
Z.~Khasidashvili, M.~Kinanah, and A.~Voronkov.
\newblock Verifying equivalence of memories using a first order logic theorem
  prover.
\newblock In {\em Proceedings of 9th International Conference on Formal Methods
  in Computer-Aided Design, {FMCAD} 2009, 15-18 November 2009, Austin, Texas,
  {USA}}, 2009, pp. 128--135. {IEEE}.

\bibitem{iProver08}
K.~Korovin.
\newblock iProver - An Instantiation-Based Theorem Prover for First-Order Logic
  (System Description).
\newblock In Armando et~al. \cite{DBLP:conf/cade/2008}, pp. 292--298.

\bibitem{InstaGen13}
K.~Korovin.
\newblock Inst-Gen - A Modular Approach to Instantiation-Based Automated
  Reasoning.
\newblock In Voronkov and Weidenbach \cite{DBLP:conf/birthday/2013ganzinger},
  pp. 239--270.

\bibitem{Noncyclic13}
K.~Korovin.
\newblock Non-cyclic Sorts for First-Order Satisfiability.
\newblock In P.~Fontaine, C.~Ringeissen, and R.~A. Schmidt, eds., {\em FroCos},
  2013, {\em LNCS 8152}, pp. 214--228. Springer.

\bibitem{LassezM87}
J.-L. Lassez and K.~Marriott.
\newblock Explicit Representation of Terms Defined by Counter Examples.
\newblock {\em J. Autom. Reasoning}, 3(3):301--317, 1987.

\bibitem{Lewis80}
H.~R. Lewis.
\newblock Complexity Results for Classes of Quantificational Formulas.
\newblock {\em J. Comput. Syst. Sci.}, 21(3):317--353, 1980.

\bibitem{Mace4Ref03}
W.~McCune.
\newblock Mace4 Reference Manual and Guide.
\newblock {\em CoRR}, cs.SC/0310055, 2003.

\bibitem{NieuwenhuisOT04}
R.~Nieuwenhuis, A.~Oliveras, and C.~Tinelli.
\newblock Abstract {DPLL} and Abstract {DPLL} Modulo Theories.
\newblock In F.~Baader and A.~Voronkov, eds., {\em Logic for Programming,
  Artificial Intelligence, and Reasoning, 11th International Conference, {LPAR}
  2004, Montevideo, Uruguay, March 14-18, 2005, Proceedings}, 2004, {\em LNCS
  3452}, pp. 36--50. Springer.

\bibitem{NieuwenhuisR01}
R.~Nieuwenhuis and A.~Rubio.
\newblock Paramodulation-Based Theorem Proving.
\newblock In Robinson and Voronkov \cite{DBLP:books/el/RobinsonV01}, pp.
  371--443.

\bibitem{GeoRes06}
H.~de~Nivelle and J.~Meng.
\newblock Geometric Resolution: A Proof Procedure Based on Finite Model Search.
\newblock In U.~Furbach and N.~Shankar, eds., {\em IJCAR}, 2006, {\em LNCS
  4130}, pp. 303--317. Springer.

\bibitem{PerezV07}
J.~A.~N. P{\'{e}}rez and A.~Voronkov.
\newblock Encodings of Bounded {LTL} Model Checking in Effectively
  Propositional Logic.
\newblock In F.~Pfenning, ed., {\em Automated Deduction - CADE-21, 21st
  International Conference on Automated Deduction, Bremen, Germany, July 17-20,
  2007, Proceedings}, 2007, {\em LNCS 4603}, pp. 346--361. Springer.

\bibitem{Generalization08}
J.~A.~N. P{\'e}rez and A.~Voronkov.
\newblock Proof Systems for Effectively Propositional Logic.
\newblock In Armando et~al. \cite{DBLP:conf/cade/2008}, pp. 426--440.

\bibitem{PerezV13}
J.~A.~N. P{\'{e}}rez and A.~Voronkov.
\newblock Planning with Effectively Propositional Logic.
\newblock In Voronkov and Weidenbach \cite{DBLP:conf/birthday/2013ganzinger},
  pp. 302--316.

\bibitem{Pichler03}
R.~Pichler.
\newblock Explicit versus implicit representations of subsets of the Herbrand
  universe.
\newblock {\em Theor. Comput. Sci.}, 290(1):1021--1056, 2003.

\bibitem{DPLLSX10}
R.~Piskac, L.~M. de~Moura, and N.~Bj{\o}rner.
\newblock Deciding Effectively Propositional Logic Using DPLL and Substitution
  Sets.
\newblock {\em J. Autom. Reasoning}, 44(4):401--424, 2010.

\bibitem{DBLP:books/el/RobinsonV01}
J.~A. Robinson and A.~Voronkov, eds.
\newblock {\em Handbook of Automated Reasoning (in 2 volumes)}.
\newblock Elsevier and {MIT} Press, 2001.

\bibitem{SilvaS96}
J.~P.~M. Silva and K.~A. Sakallah.
\newblock Conflict Analysis in Search Algorithms for Satisfiability.
\newblock In {\em ICTAI}, 1996, pp. 467--469.

\bibitem{FinderRef94}
J.~K. Slaney.
\newblock FINDER: Finite Domain Enumerator - System Description.
\newblock In A.~Bundy, ed., {\em CADE}, 1994, {\em LNCS 814}, pp. 798--801.
  Springer.

\bibitem{SudaWeidenbachWischnewskiIJCAR10}
M.~Suda, C.~Weidenbach, and P.~Wischnewski.
\newblock {On the Saturation of YAGO}.
\newblock In J.~Giesl and R.~H\"ahnle, eds., {\em {Automated Reasoning}},
  Edinburgh, UK, 2010, {\em {Lecture Notes in Artificial Intelligence}}, vol.
  6173, pp. 441--456. Springer.

\bibitem{Tammet03finitemodel}
T.~Tammet.
\newblock Finite Model Building: Improvements and Comparisons.
\newblock In {\em In: Model Computation � Principles, Algorithms,
  Applications, CADE-19 Workshop W4}, 2003.

\bibitem{DBLP:conf/birthday/2013ganzinger}
A.~Voronkov and C.~Weidenbach, eds.
\newblock {\em Programming Logics - Essays in Memory of Harald Ganzinger},
  2013, {\em LNCS 7797}. Springer.

\bibitem{Weidenbach01}
C.~Weidenbach.
\newblock Combining Superposition, Sorts and Splitting.
\newblock In Robinson and Voronkov \cite{DBLP:books/el/RobinsonV01}, pp.
  1965--2013.

\bibitem{SEMRef95}
J.~Zhang and H.~Zhang.
\newblock SEM: a System for Enumerating Models.
\newblock In {\em IJCAI}, 1995, pp. 298--303. Morgan Kaufmann.

\end{thebibliography}
